\documentclass[twoside,twocolumn,9pt]{article}
\usepackage{extsizes}
\usepackage[super,sort&compress,comma]{natbib} 
\usepackage[version=3]{mhchem}
\usepackage[left=1.5cm, right=1.5cm, top=1.785cm, bottom=2.0cm]{geometry}
\usepackage{balance}
\usepackage{mathptmx}
\usepackage{sectsty}
\usepackage{graphicx} 
\usepackage{lastpage}
\usepackage[format=plain,justification=justified,singlelinecheck=false,font={stretch=1.125,small,sf},labelfont=bf,labelsep=space]{caption}
\usepackage{float}
\usepackage{fancyhdr}
\usepackage{fnpos}
\usepackage[english]{babel}
\addto{\captionsenglish}{%
  \renewcommand{\refname}{Notes and references}
}
\usepackage{array}
\usepackage{droidsans}
\usepackage{charter}
\usepackage[T1]{fontenc}
\usepackage[usenames,dvipsnames]{xcolor}
\usepackage{setspace}
\usepackage[compact]{titlesec}
\usepackage{hyperref}
\usepackage{amsmath}
\usepackage{siunitx}
\usepackage{amssymb}
\usepackage{pdfpages}

\definecolor{cream}{RGB}{222,217,201}

\begin{document}

\pagestyle{fancy}
\thispagestyle{plain}
\fancypagestyle{plain}{
%%%HEADER%%%
\renewcommand{\headrulewidth}{0pt}
}
%%%END OF HEADER%%%

%%%PAGE SETUP - Please do not change any commands within this section%%%
\makeFNbottom
\makeatletter
\renewcommand\LARGE{\@setfontsize\LARGE{15pt}{17}}
\renewcommand\Large{\@setfontsize\Large{12pt}{14}}
\renewcommand\large{\@setfontsize\large{10pt}{12}}
\renewcommand\footnotesize{\@setfontsize\footnotesize{7pt}{10}}
\makeatother

\renewcommand{\thefootnote}{\fnsymbol{footnote}}
\renewcommand\footnoterule{\vspace*{1pt}% 
\color{cream}\hrule width 3.5in height 0.4pt \color{black}\vspace*{5pt}} 
\setcounter{secnumdepth}{5}

\makeatletter 
\renewcommand\@biblabel[1]{#1}            
\renewcommand\@makefntext[1]% 
{\noindent\makebox[0pt][r]{\@thefnmark\,}#1}
\makeatother 
\renewcommand{\figurename}{\small{Fig.}~}
\sectionfont{\sffamily\Large}
\subsectionfont{\normalsize}
\subsubsectionfont{\bf}
\setstretch{1.125} %In particular, please do not alter this line.
\setlength{\skip\footins}{0.8cm}
\setlength{\footnotesep}{0.25cm}
\setlength{\jot}{10pt}
\titlespacing*{\section}{0pt}{4pt}{4pt}
\titlespacing*{\subsection}{0pt}{15pt}{1pt}
%%%END OF PAGE SETUP%%%

%%%FOOTER%%%
\fancyfoot{}
\fancyfoot[LO,RE]{\vspace{-7.1pt}\includegraphics[height=9pt]{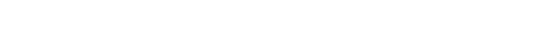}}
\fancyfoot[CO]{\vspace{-7.1pt}\hspace{13.2cm}\includegraphics{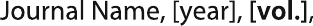}}
\fancyfoot[CE]{\vspace{-7.2pt}\hspace{-14.2cm}\includegraphics{head_foot/RF}}
\fancyfoot[RO]{\footnotesize{\sffamily{1--\pageref{LastPage} ~\textbar  \hspace{2pt}\thepage}}}
\fancyfoot[LE]{\footnotesize{\sffamily{\thepage~\textbar\hspace{3.45cm} 1--\pageref{LastPage}}}}
\fancyhead{}
\renewcommand{\headrulewidth}{0pt} 
\renewcommand{\footrulewidth}{0pt}
\setlength{\arrayrulewidth}{1pt}
\setlength{\columnsep}{6.5mm}
\setlength\bibsep{1pt}
%%%END OF FOOTER%%%

%%%FIGURE SETUP - please do not change any commands within this section%%%
\makeatletter 
\newlength{\figrulesep} 
\setlength{\figrulesep}{0.5\textfloatsep} 

\newcommand{\topfigrule}{\vspace*{-1pt}% 
\noindent{\color{cream}\rule[-\figrulesep]{\columnwidth}{1.5pt}} }

\newcommand{\botfigrule}{\vspace*{-2pt}% 
\noindent{\color{cream}\rule[\figrulesep]{\columnwidth}{1.5pt}} }

\newcommand{\dblfigrule}{\vspace*{-1pt}% 
\noindent{\color{cream}\rule[-\figrulesep]{\textwidth}{1.5pt}} }

\makeatother
%%%END OF FIGURE SETUP%%%

%%%TITLE, AUTHORS AND ABSTRACT%%%
\twocolumn[
  \begin{@twocolumnfalse}
{\includegraphics[height=30pt]{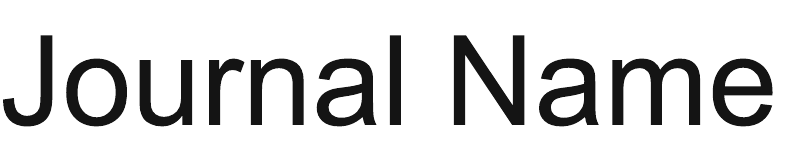}\hfill\raisebox{0pt}[0pt][0pt]{\includegraphics[height=55pt]{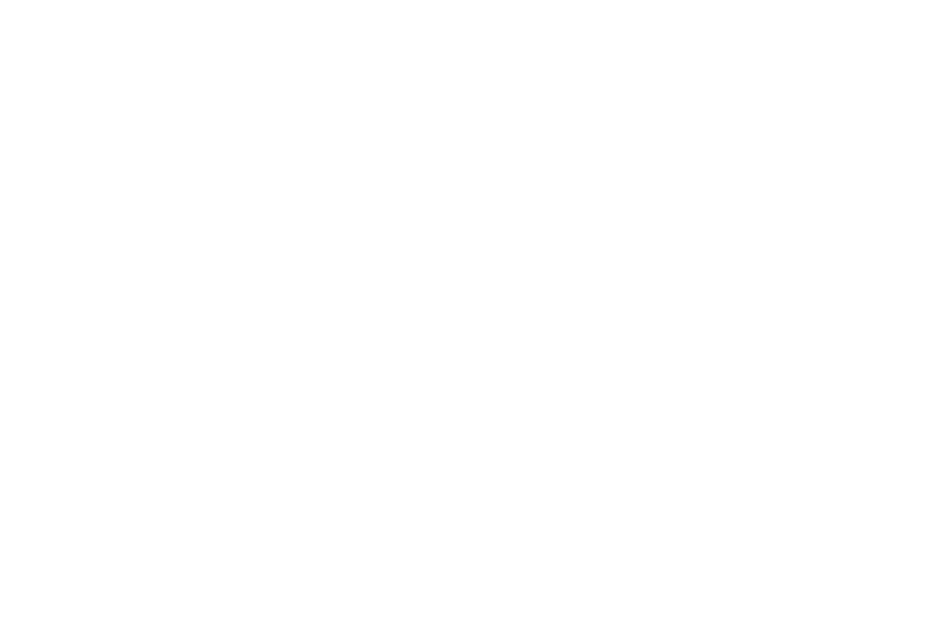}}\\[1ex]
\includegraphics[width=18.5cm]{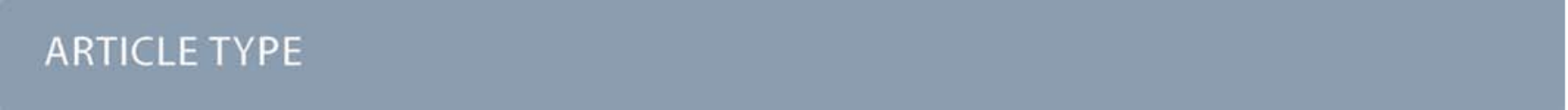}}\par
\vspace{1em}
\sffamily
\begin{tabular}{m{4.5cm} p{13.5cm} }

\includegraphics{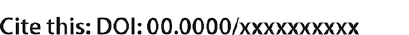} & \noindent\LARGE{\textbf{Avidity and surface mobility in multivalent ligand-receptor binding
$^\dag$}} \\
\vspace{0.3cm} & \vspace{0.3cm} \\

 & \noindent\large{Simon Merminod,\textit{$^{a\,\ddag}$} John~R. Edison,\textit{$^{b\,\P}$}, Huang Fang,\textit{$^{a}$} Michael~F. Hagan,\textit{$^{a}$} and W.~Benjamin Rogers\textit{$^{{\ast}a}$}} \\

\includegraphics{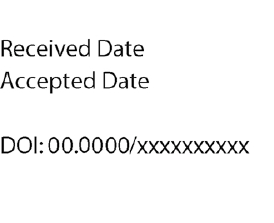} & \noindent\normalsize{Targeted drug delivery relies on two physical processes: the selective binding of a therapeutic particle to receptors on a specific cell membrane, followed by transport of the particle across the membrane. 
In this article, we address some of the challenges in controlling the thermodynamics and dynamics of these two processes by combining a simple experimental system with a statistical mechanical model. 
Specifically, we characterize and model multivalent ligand-receptor binding between colloidal particles and fluid lipid bilayers, as well as the surface mobility of membrane-bound particles. 
We show that the mobility of the receptors within the fluid membrane is key to both the thermodynamics and dynamics of binding.
First, we find that the particle-membrane binding free energy---or avidity---is a strongly nonlinear function of the ligand-receptor affinity. We attribute the nonlinearity to a combination of multivalency and recruitment of fluid receptors to the binding site. Our results also suggest that partial wrapping of the bound particles by the membrane enhances avidity further. 
Second, we demonstrate that the lateral mobility of membrane-bound particles is also strongly influenced by the recruitment of receptors. Specifically, we find that the lateral diffusion coefficient of a membrane-bound particle is dominated by the hydrodynamic drag against the aggregate of receptors within the membrane. 
These results provide one of the first direct validations of the working theoretical framework for multivalent interactions.
They also highlight that the fluidity and elasticity of the membrane are as important as the the ligand-receptor affinity in determining the binding and transport of small particles attached to membranes.}
% {The abstract should be a single paragraph which summarises the content of the article. Any references in the abstract should be written out in full \textit{e.g.}\ [Surname \textit{et al., Journal Title}, 2000, \textbf{35}, 3523].}\\%The abstrast goes here instead of the text "The abstract should be..."

\end{tabular}

 \end{@twocolumnfalse} \vspace{0.6cm}

  ]
%%%END OF TITLE, AUTHORS AND ABSTRACT%%%

%%%FONT SETUP - please do not change any commands within this section
\renewcommand*\rmdefault{bch}\normalfont\upshape
\rmfamily
\section*{}
\vspace{-1cm}

%%%FOOTNOTES%%%

\footnotetext{\textit{$^{\ast}$~To whom correspondence should be addressed. E-mail: wrogers@brandeis.edu.}}
\footnotetext{\textit{$^{a}$~Martin~A. Fisher School of Physics, Brandeis University, Waltham, MA~02453, USA.}}
\footnotetext{\textit{$^{b}$~Brandeis Materials Research Science and Engineering Center, Brandeis University, Waltham, MA~02453, USA.}}
\footnotetext{\textit{$^{\ddag}$~Present address: Department of Molecular and Cellular Biology, Harvard University, Cambridge, MA~02138, USA.}}
\footnotetext{\textit{$^{\P}$~Present address: Department of Chemical and Biomolecular Engineering, Johns Hopkins University, Baltimore, MD~21218, USA.}}

%Please use \dag to cite the ESI in the main text of the article.
\footnotetext{\dag~Electronic Supplementary Information (ESI) available: Text and figures about the DNA constructs, the colloidal particle synthesis and ligand grafting, the sample preparation, fluorescence recovery after photobleaching, the optical setup, the calibration of the total internal reflection microscope, the interaction potentials, the avidity, the theory of multivalent ligand-receptor binding, our model and numerical scheme, receptor recruitment, particle wrapping by the membrane, and the mobility of membrane-bound particles. See DOI: 00.0000/00000000.}

%%%END OF FOOTNOTES%%%

\begin{figure*}[t]
    \centering
    \includegraphics{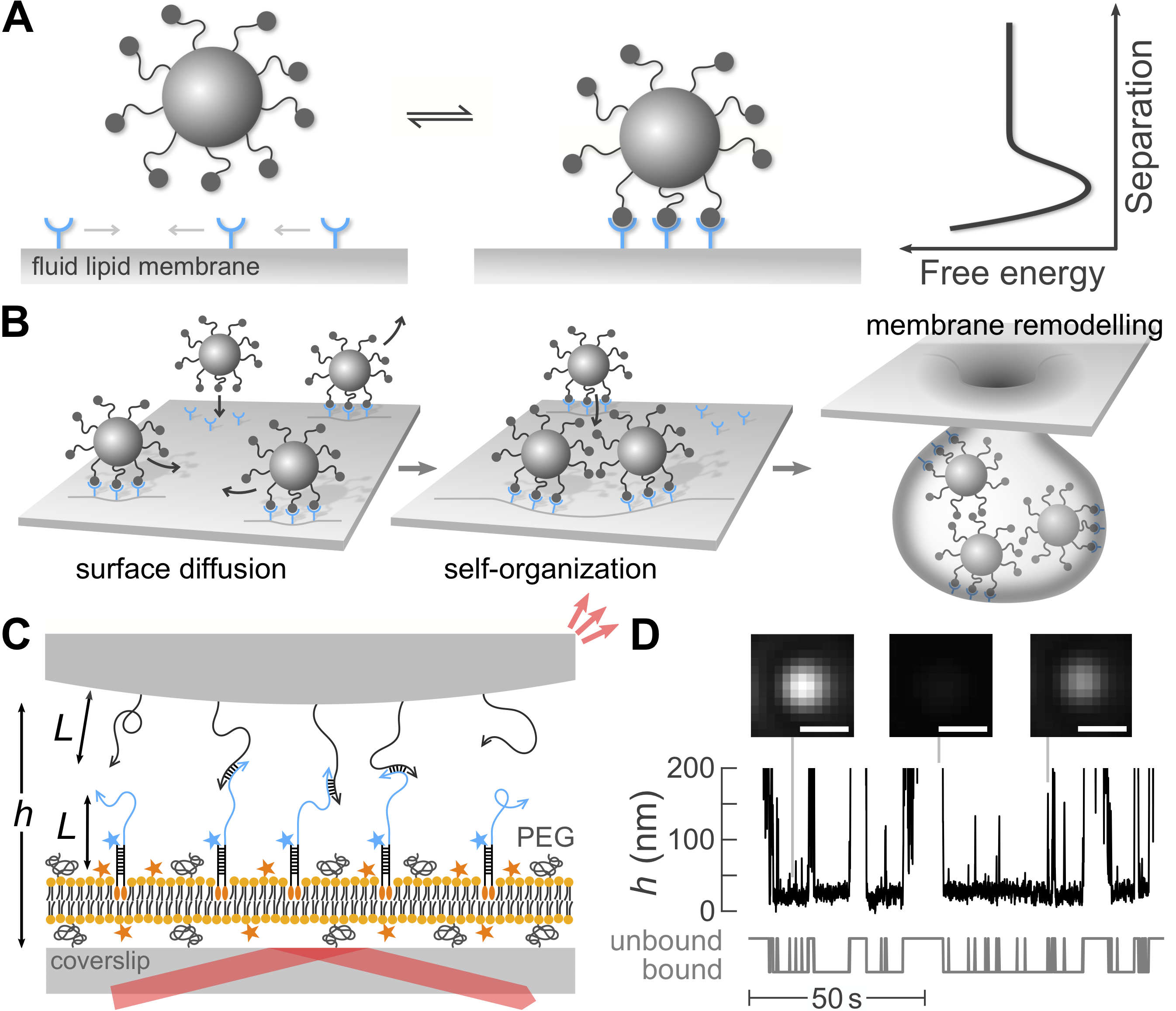}
    \caption{
    Overview of the coupled physical processes that we study and our experimental system.
    (A)~
    Transient, multivalent ligand-receptor binding gives rise to an effective interaction potential between a colloidal particle and a fluid lipid membrane. How does the interaction potential depend on the ligand-receptor affinity, as well as the fluidity and elasticity of the membrane?
    (B)~
    The lateral mobility of membrane-bound particles can enable membrane remodelling and cellular uptake. How does the mobility depend on the details of the particle-membrane interactions?
    (C)~
    In our experimental system, 
    DNA ligands grafted to a 1.4-\si{{\micro}m}-diameter colloidal particle hybridize with DNA receptors embedded in a supported lipid bilayer (SLB) to induce attractive interactions. 
    Single-stranded DNA ligands and receptors have a typical length $L \approx 15$~nm. 
    DNA receptors are labelled with 6-FAM fluorophores (blue stars) and anchored in the membrane using double cholesterol-triethylene glycol (TEG) modifications (orange ovals).
    The SLB is formed of a mixture of three phospholipid species, including 2.4\%~(w/w) that are labelled with Texas Red fluorophores (orange stars) and 0.5\%~(w/w) that are \mbox{PEG}ylated (black coils). 
    The particle scatters light (small red arrows) from an evanescent wave generated by a reflected laser beam (large red arrow).
    (D)~The inferred colloid-glass separation, \textit{h}, shows intermittent binding. A colloid is considered bound when $h$ is smaller than a threshold value, roughly 45~nm. Insets show micrographs of the light scattered by the single colloidal particle at different moments along the trajectory. Bright spots correspond to small separations. Scalebars, 1~\si{{\micro}m}.}
    \label{fig:fig1}
\end{figure*}

%%%MAIN TEXT%%%%
Achieving the targeted binding of small particles to cell membranes has the potential to improve strategies for drug delivery; yet designing such interactions is challenging because the interactions are multivalent and the membranes are fluid and deformable.
The basic idea is to coat a therapeutic payload with specific molecular species that bind it selectively---ideally exclusively---to cells of a specific identity, thereby maximizing a drug's efficacy while minimizing toxic side effects.\cite{ParkJcontrolRelease2011, TorchilinNatRevDrugDiscov2014, ZhangAdvDrugDeliverRev2010, WilhelmNatRevMat2016, TiwariNanomedicine2010} Achieving targeted binding thus requires the ability to design or select ligands that can recognize the biochemical attributes of the target cell without also binding to off-target membranes. This task is challenging because cell membranes comprise complex collections of various receptors with different concentrations depending on the cell's identity and health.\cite{MolBiolCellBook} What is the optimal strategy for picking out one membrane composition over all the rest? Answering this question is complicated by the fact that the interactions between particles and membranes are typically multivalent---multiple pairs of ligands and receptors interact simultaneously---and the receptors in cell membranes are typically mobile and can diffuse on the membrane surface \cite{FrenkelPNAS2011, FrenkelPNAS2017, KiesslingACSChemBiol2007} (Fig.~\ref{fig:fig1}A). Therefore, while chemical complementarity ensures specific recognition between individual ligands and receptors, the selectivity of the binding response to molecular recognition is much more complex.

Simple in vitro systems and theoretical models provide a path toward a better fundamental understanding of multivalent interactions and how to design them. Qualitatively, the strength and specificity of multivalent interactions result from a subtle balance between enthalpy and entropy.\cite{WhitesidesAngewChem1998} In the last decade, much progress has been made in developing theoretical models to predict multivalent interactions by combining the tools of statistical physics with other system-specific frameworks, such as theories of membrane mechanics. This approach has been applied to situations in which the ligands and receptors are fixed \cite{FrenkelJChemPhys2012, FrenkelPCCP2016} or mobile,\cite{FrenkelPRL2014} and the interfaces are rigid or deformable.\cite{DiMichelePCCP2015,MognettiPRE2018,DiMicheleRepProgPhys2019}  However, experimental validations of these various models lag behind.\cite{DiMicheleRepProgPhys2019} While a few recent experimental studies have explored the physical mechanisms underlying the kinetics of binding, the membrane deformation due to multivalent interactions, or the factors influencing binding selectivity,\cite{HookNanoLett2016,DiMicheleNatCommun2015,DiMichelePCCP2015,MognettiSoftMatter2016,BruylantsLangmuir2019,Scheepers2020,Liu2020} they do not measure the particle-membrane binding free energy (Fig.~\ref{fig:fig1}A). Therefore, direct experimental measurements of the binding free energy are needed to first validate, and then design, specific interactions between particles and membranes.

In addition to specific binding, the lateral mobility of membrane-bound particles also plays a major role in transport of a therapeutic payload across a membrane. While a single bound particle whose adhesion energy is large compared with the bending rigidity of the membrane might be able to cross the membrane alone, by becoming fully wrapped by the membrane, more weakly bound particles cannot.\cite{DesernoPRE2004} Instead, multiple weakly bound particles have to first self-organize via lateral attractions \cite{SafinyaPRL1999,DesernoNature2007, DufresnePRE2016, KraftSciRep2016} to trigger remodelling of the membrane, such as collective budding \cite{DesernoNature2007, CacciutoSoftMatter2013, DinsmoreNanoscale2019} (Fig.~\ref{fig:fig1}B), in order to enter the cell.
The success---or failure---of such processes hinges on surface mobility: For weakly bound particles to self-organize, they must find one another and sample configurations within their energy landscape faster than they unbind from the membrane. Thus, to engineer successful strategies for targeted delivery, we also need to understand how the lateral mobility of membrane-bound particles depends on the details of ligand-receptor binding.

In this article, we combine experiments and theory to characterize the relationships between the ligand-receptor affinity, the binding avidity, and the lateral mobility of colloidal particles interacting with supported bilayer membranes. We use complementary single-stranded DNA molecules as model ligand-receptor pairs, and  measure the emergent interactions using total internal reflection microscopy. Using DNA as ligands and receptors is crucial since it enables us to precisely tune the affinity in situ by adjusting the temperature. We find that the mobility of the receptors within the membrane plays a key role in determining both the avidity and the particle mobility, highlighting the importance of membrane fluidity in targeted delivery. Specifically, we find that the avidity is a strongly nonlinear function of the ligand-receptor affinity. A statistical mechanical model of the interactions shows that this nonlinear dependence results from a combination of multivalency and recruitment of receptors to the site of contact between the particle and the membrane. Disagreements between our measurements and model predictions in the limit of strong binding suggest that elastic membrane deformations further enhance the nonlinearity of avidity, by bending the membrane to increase the area of contact between the particle and the membrane. Combining measurements of the lateral diffusion of membrane-bound particles with predictions of the number and spatial distribution of bound receptors, we also show that the diffusion coefficient of membrane-bound particles is determined by the hydrodynamic drag against the aggregate of recruited receptors, and not by the viscosity of the surrounding solution. Taken together, our results show that the avidity and surface mobility of particles interacting with fluid membranes---two key ingredients in targeted delivery---are related through the mobility of the receptors. Therefore, our findings suggest that future attempts to design interactions to target specific cell membranes should consider the membrane fluidity and elasticity, in addition to the composition of receptors expressed on the membrane surface.

\section*{Results and Discussion}

Our experimental system consists of DNA-coated colloidal particles and DNA-functionalized supported lipid bilayers (SLBs). The particles are 1.4-\si{{\micro}m}-diameter spheres made of 3-(trimethoxysilyl) propyl methacrylate (TPM), which are coated with single-stranded DNA oligomers using click chemistry \cite{PineNatCommun2015} (Fig.~\ref{fig:fig1}C) . The supported lipid bilayers are comprised of 97.1\%~(w/w) 1,2-dioleoyl-sn-glycero-3-phosphocholine (18:1~DOPC), 2.4\%~(w/w) PEG(2k)-labeled 1,2-dioleoyl-sn-glycero-3-phosphoethanolamine (18:1~PE), and 0.5\%~(w/w) Texas Red-labeled 1,2-dihexadecanoyl-sn-glycero-3-phosphoethanolamine (DHPE). We make the supported bilayers by spreading liposomes on a cleaned glass coverslip. After spreading, we label the supported bilayer with DNA using a double-stranded DNA handle modified with two cholesterol molecules. One of the DNA handles is also modified with the fluorophore FAM. The PEGylated lipids ensure mobility of the receptors within the supported bilayer and prevent nonspecific binding of the particles to the membrane. The Texas Red-labeled lipids and FAM-labeled DNA molecules allow us to image the SLB and DNA coatings, and measure their fluidity. We verify that the lipids and surface-anchored DNA strands are mobile using fluorescence recovery after photobleaching (FRAP). See Materials and Methods, as well as the ESI$^{\dag}$ for further details.

We measure the interactions and the lateral diffusion of colloidal particles from their three-dimensional trajectories, using total internal reflection microscopy. Briefly, a laser beam totally internally reflected from a glass-water interface creates an evanescent wave in the sample chamber (Fig.~\ref{fig:fig1}C). A colloidal particle within the evanescent wave scatters an amount of light that decreases exponentially with the separation between the particle and the coverslip $h$.\cite{Prieve1999} We image the scattered light onto a \mbox{sCMOS} camera and quantify the scattered intensity, $I(t)$, using existing particle-tracking routines.\cite{Crocker1996} Using a calibration based on the hydrodynamic coupling of a sphere and a flat interface,\cite{Prieve1999,VolpeOptExp2009} we infer the vertical position of the particle as a function of time, $h(t)$, from the scattered intensity, $I(t)$. We record videos at 100~Hz for a duration of 500~s, and image an average of 5 particles simultaneously. See the ESI$^{\dag}$ for a detailed description of the experimental setup and the calibration method.

\begin{figure*}[t]
    \centering
    \includegraphics{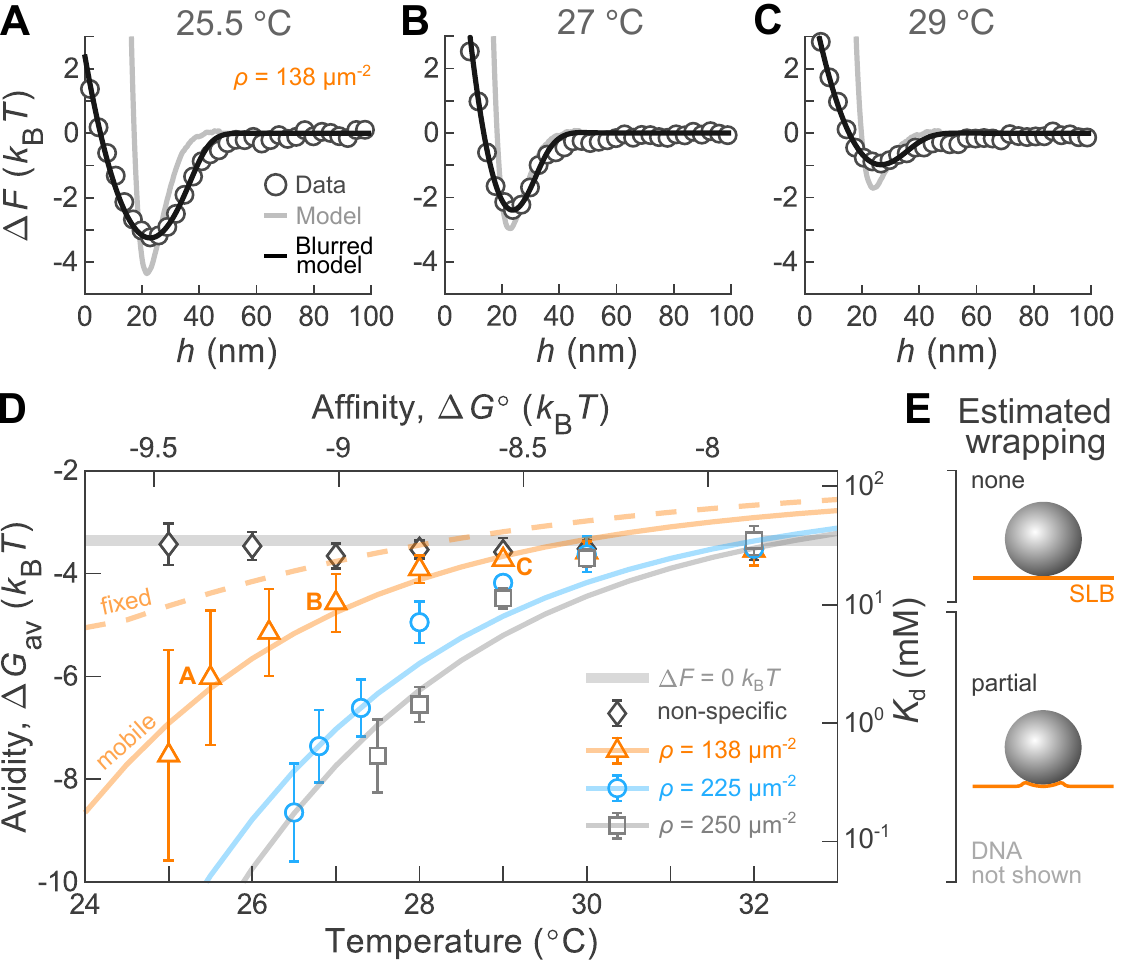}
    \caption{ Colloid-membrane interactions.
    (A--C)~Example interaction potentials at fixed receptor density $\rho = 138$~\si{{\micro}m^{-2}} and increasing temperature show a temperature-dependent attractive well for $T=25.5$~\si{\degreeCelsius} (A), 27~\si{\degreeCelsius} (B), and 29~\si{\degreeCelsius} (C). Circles are experimental data.
    Gray curves are the model; black curves are the model after blurring to account for the finite precision of our measurements. (D)~The avidity, $\Delta G_\textrm{av}$, defined in the text (Eq.~\ref{eq:avidity}), as a function of temperature and affinity, $\Delta G^\circ$, for three receptor densities: $\rho = 138/\si{{\micro}m^2}$ (orange), $\rho = 225/\si{{\micro}m^2}$ (blue), and $\rho = 250/\si{{\micro}m^2}$ (gray). Experimental measurements (symbols) show that avidity is a strongly non-linear function of temperature. Solid curves are best-fit model predictions at fixed receptor density. Error bars denote the standard deviation of avidities measured for multiple particles. Annotations ``A'', ``B'' and ``C'' correspond to the conditions of panels~A, B and~C. The dashed orange curve shows avidity from a model with \textit{fixed} receptors, in contrast to the  solid orange curve  which shows predictions with  \textit{mobile} receptors. Experimental measurements of particles coated with noncomplementary ligands (diamonds) show a weak temperature-independent, nonspecific attraction. The thick gray curve shows the avidity for $\Delta F(h)=0$ for all separations. The right axis shows the dissociation constant, $K_\textrm{d} = c_\circ\,\textrm{e}^{\,\Delta G_\textrm{av}/k_\textrm{B}T}$, with $c_\circ = 1$~mol/l.
    (E)~We estimate that partial wrapping of the particle by the membrane occurs for avidities stronger than $-4.7~k_\textrm{B}T$, or $K_\textrm{d} < 10$~mM.
    }
    \label{fig:fig2-avidity}
\end{figure*}

\subsection*{Emergent colloidal-membrane interactions.}

\paragraph*{Experimental measurements.}

We infer the colloid-membrane interaction potentials from the separation time series for each particle.
Fig.~\ref{fig:fig1}D shows an example time series, in which the particle intermittently binds to and unbinds from the supported bilayer. Assuming that the particle is in thermal equilibrium, the distribution of its vertical separation $h$ obeys Boltzmann statistics. Therefore, we measure the interaction potential between the particle and the membrane, $\Delta F_\textrm{tot}(h)$, up to a constant by creating a histogram of the separations, $P(h)$, and then inverting the Boltzmann distribution, $P(h)~\propto~\exp\,[-\Delta F_\textrm{tot}(h)/k_\textrm{B} T]$, where $k_\textrm{B}$ is the Boltzmann constant and $T$ is the temperature. Finally, we subtract the linear contribution to $\Delta F_\textrm{tot}(h)$ due to gravity to obtain the DNA-mediated interaction potentials, $\Delta F(h)$. We measure ${\Delta}F(h)$ for a number of different temperatures and three different receptor densities.

The interaction potentials that we measure feature a short-range attractive well whose depth depends sensitively on temperature. Fig.~\ref{fig:fig2-avidity}A--C show examples of three interaction potentials for three different temperatures. All interaction potentials show a very short-range repulsion and a short-range attraction, with a well depth that increases with decreasing temperature. The range of the attractive wells spans roughly 20--30~nm above the membrane, which is comparable to twice the end-to-end distance of the grafted DNA molecules.\cite{Rogers2011}

We compute the avidity, $\Delta G_\textrm{av}$ for each interaction potential.
Whereas the affinity tells us about the free energy of binding between a single ligand-receptor pair, the avidity tells us about the free energy of binding between the particle and membrane, which in general involves the cooperative interactions between many ligands and receptors. To account for both the range and depth of the attractive well, we define avidity from the integral of the Boltzmann weight over the bound state,\cite{BenTal2000,HaganJChemPhys2009}
\begin{equation}
    \Delta G_\textrm{av} = -k_\textrm{B}T\,\log\left[ (c_\circ N_\textrm{A})^{1/3} \int_0^{\lambda_\textrm{b}} e^{-{\Delta}F(h)/k_\textrm{B}T} \textrm{d}h \right],
    \label{eq:avidity}
\end{equation}
where $c_\circ = 1$~mol/l is a reference concentration, $N_\textrm{A}$ is Avogadro's number, and $\lambda_\textrm{b}$ is the maximum separation within which ligands and receptors can bind. We set $\lambda_\textrm{b} = 34$~nm for all calculations (see the ESI$^{\dag}$ for details). Finally, we note that while avidity is a negative number, throughout the discussion, we use ``increasing avidity'' to mean a more negative value of avidity, and thus stronger binding.

We find that the avidity is a strongly nonlinear function of temperature, and increases with receptor density. Fig.~\ref{fig:fig2-avidity}D shows the experimentally measured avidities as a function of temperature for three receptor densities. 
In all three cases, the avidity decreases upon increasing temperature and increases with increasing receptor density. Furthermore, the avidity becomes a sharper function of temperature upon increasing receptor density, decreasing from $-8~k_\textrm{B}T$ to $-3.5~k_\textrm{B}T$ over roughly five degrees Celsius at the highest density. The thick gray line in Fig.~\ref{fig:fig2-avidity}D is the avidity for a system with $\Delta F =0$ over the entire binding region, namely $\Delta G_\textrm{av} = - k_\textrm{B}T\,\log\left[ \lambda_\textrm{b}(c_\circ N_\textrm{A})^{1/3}\right]$, which arises from the probability of particles residing within the binding volume even in the absence of interactions. 
Control experiments using particles grafted with noncomplementary DNA sequences yield an avidity of roughly $-3.5~k_\textrm{B}T$ at all temperatures. 
While this value is close to the avidity for $\Delta F=0$, we note that particles are excluded from the region of small separation, $h \lesssim 20$~nm, and thus this value implies a weak attraction for separations between 20~nm and $\lambda_\textrm{b}$.
We speculate that this weak nonspecific attraction results from nonspecific interactions between the TPM particles and the PEG molecules grafted to the membrane.
Finally, although we present our experimental findings in terms of their relationship to the temperature, we note that the key physical quantity is actually the ligand-receptor affinity, which itself is a linear function of the temperature.\cite{SantaLuciaPNAS1998}

These qualitative relationships between avidity, affinity, and receptor density are consistent with physical intuition---greater affinity and larger receptor densities favor the formation of ligand-receptor bonds. However, unpacking the emergent interactions further requires a theoretical model. In the following, we use a statistical mechanical model and compare its predictions to our experimental measurements. The advantage of such a microscopic model is that it allows us to dissect the relevant contributors to the avidity, such as multivalency, receptor mobility, and even elastic membrane deformations. Which of these effects are dominant? And how do they alter the relationship between affinity and avidity?

\paragraph*{Theoretical model.}

The emergent interactions between two surfaces decorated with DNA molecules arise from a rich interplay of enthalpic and entropic effects. When the surfaces are close enough, complementary strands can hybridize to form bridges by Watson-Crick base pairing, lowering the enthalpy of the system. However, when two complementary strands bind, they must also incur a distance-dependent entropic penalty, as they sacrifice degrees of freedom in order to hybridize. Even the unhybridized ligands and receptors can lose configurational entropy when they are squeezed between the two surfaces, an effect which again depends on the separation distance. Finally, the mobility of the membrane-anchored receptors complicates the situation further, since the mobile receptors can enrich or deplete the confined region between the particle and the membrane by paying an associated cost in mixing entropy.\cite{DiMicheleRepProgPhys2019}

We model the particle-membrane interactions using a statistical mechanical theory of multivalent interactions developed by Mognetti, Frenkel, and coworkers.\cite{FrenkelJChemPhys2012, DiMicheleRepProgPhys2019} We model the ligands and receptors as ideal chains with sticky ends, and the SLB as a flat plate in contact with a grand canonical reservoir of receptors. Details about the theoretical framework and our semi-analytical approach to estimate the effective particle-membrane interactions are in the ESI.$^{\dag}$ All of the model parameters are constrained by experimental measurements with the exception of the receptor grafting density, $\rho$, which we choose to obtain the best match between the modeled and experimentally measured avidities (see the ESI$^{\dag}$).

\paragraph*{Comparing experiments and theory.}

Predictions from our model 
% with mobile DNA receptors 
reproduce many aspects of our experimental measurements, with the receptor density as the only adjustable parameter.
For the lowest receptor density, predictions of the avidity closely match our experimental measurements (Fig.~\ref{fig:fig2-avidity}D). Furthermore, the model interaction potentials---convolved with a Gaussian kernel to simulate the finite precision of our measurements---reproduce the full shape of the experimental potentials (Fig.~\ref{fig:fig2-avidity}\mbox{A--C}).
For the two larger densities, the model also agrees well with experimental results, although not as closely as for the lowest density.
The only significant disagreement between theory and experiment is
the extent of the nonlinearity of avidity with respect to temperature: The avidity increases more sharply upon decreasing temperature in the experiments as compared to the model. Another minor disparity concerns the value of the plateau of avidity at high temperatures, which is lower in the experiments than in the model. This disparity arises because the experimental potentials have a small attractive well from non-specific attraction even at the highest temperatures, while the simulated potentials do not.

Importantly, the best-fit receptor densities that we find are  consistent with our experimental conditions. The values range from roughly 130--250 molecules per \si{{\micro}m^2}, corresponding to an average spacing between receptors of about 60--90~nm. These typical distances are not so small as to be incompatible with the spontaneous adsorption of receptors to the membrane. They are also not so large as to prevent the formation of multiple ligand-receptor pairs between the particle and the membrane in a reasonable amount of time. Thus the receptor densities fall within a range that is consistent with both spontaneous adsorption and multivalent binding. Finally, we note that the fitted receptor densities are all smaller than the ligand density, which is roughly 1200 ligands per \si{{\micro}m^2}. As a result, there are many more ligands than receptors in the gap between a particle and the membrane, which can drive recruitment of fluid receptors during binding.

We hypothesize that elastic membrane deformations---which we do not model here---explain the disagreement between theory and experiment at low temperatures. Because fluid membranes are elastic, they can deform upon binding of a particle. As shown by Deserno,\cite{DesernoPRE2004} and recently verified for nanoparticles binding to lipid vesicles,\cite{DinsmoreNanoscale2019} one can predict the onset of such deformations in free membranes from the dimensionless ratio $\Tilde{w}= 2wa^2/\kappa$, where $w$ is the particle-membrane adhesion energy per unit area, $a$ is the radius of the adhering particle, and $\kappa$ is the membrane bending rigidity. Taking $\sigma$ to be the membrane tension and defining $\Tilde{\sigma}=\sigma a^2/\kappa$, the theory predicts that the membrane will remain flat when $\Tilde{w} <4$, undergo small deformations and partial particle wrapping when $4< \Tilde{w} < 4+2\Tilde{\sigma}$, and fully wrap the particle when $\Tilde{w}> 4+2\Tilde{\sigma}$. Taking typical values for DOPC supported bilayers---$\sigma = 1$~\si{pN.nm^{-1}} and $\kappa = 20~k_\textrm{B}T$ \cite{GiannottiNanoscale2018,DinsmoreNanoscale2019}---we estimate that $\Tilde{\sigma} \approx 6000$ and $\Tilde{w} \approx 1$--9 over our range of avidities.  More specifically, we expect the membrane to remain undeformed in our system when the avidity is smaller than $-4.7~k_\textrm{B}T$, and partially wrap the particles at stronger avidities (Fig.~\ref{fig:fig2-avidity}E and more details in the ESI$^{\dag}$).

Our measurements of avidity agree with these expectations of membrane deformation. For the two largest receptor densities, we observe that the avidity is a steeper function of temperature below about $-5~k_\textrm{B}T$, which is roughly equal to our estimate of the onset of membrane bending of $-4.7~k_\textrm{B}T$ (Fig.~\ref{fig:fig2-avidity}D,E). Even though the model by Deserno concerns free---not supported---membranes,\cite{DesernoPRE2004} we believe that it is relevant to the present discussion because the membrane deformations are likely smaller than the size of our PEG spacers, which separate the membrane from the glass substrate by roughly 3~nm.\cite{MarshBBA2003} In fact, for the largest avidity that we measure, roughly $-9~k_\textrm{B}T$, we estimate that the patch of deformed membrane has a diameter of roughly 40~nm, and a deflection of only 0.5~nm above the flat membrane.\cite{DesernoPRE2004} Unfortunately, we do not have the resolution to confirm this prediction.
Finally, we suspect that membrane mechanics are modified by the addition of the DNA receptors.  In fact, we observe various membrane instabilities upon the addition of receptors, such as the spontaneous formation of tubules extending tens of micrometers into the bulk. These observations indicate that the bound receptors might facilitate membrane deformation, which is consistent with our observation that the disagreement between theory and experiment is more pronounced for the higher receptor densities.

\paragraph*{Contributions to avidity from multivalency and mobility.}

Digging further into the model, we find that both the multivalency of binding and the mobility of receptors within the fluid membrane contribute to the nonlinearity in avidity.
If the interactions were monovalent, the avidity would be a linear function of affinity. Therefore, the nonlinear dependence of avidity that we observe reflects the cooperative nature of the simultaneous interactions of many ligand-receptor pairs. 
We isolate the effect of multivalency by computing the avidity in a variation of our usual system---a simulated system in which receptors are \emph{anchored} at specific points on the membrane---but is otherwise identical to our system with \emph{mobile} receptors. In particular, the receptor grafting density in the fixed system is equal to the receptor density within the grand canonical reservoir in the mobile system, $\rho$. The difference in avidity between these two systems is thus due to receptor mobility. 
We compute the avidity in the fixed system for a single receptor density of 138~\si{{\micro}m^{-2}}.
The avidity that we obtain is again a nonlinear function of affinity, but exhibits a weaker dependence on temperature than both our experimental measurements and our predictions within the mobile system (Fig.~\ref{fig:fig2-avidity}D). 
Thus, multivalency is only one piece of the puzzle.

\begin{figure}[t!]
    \includegraphics{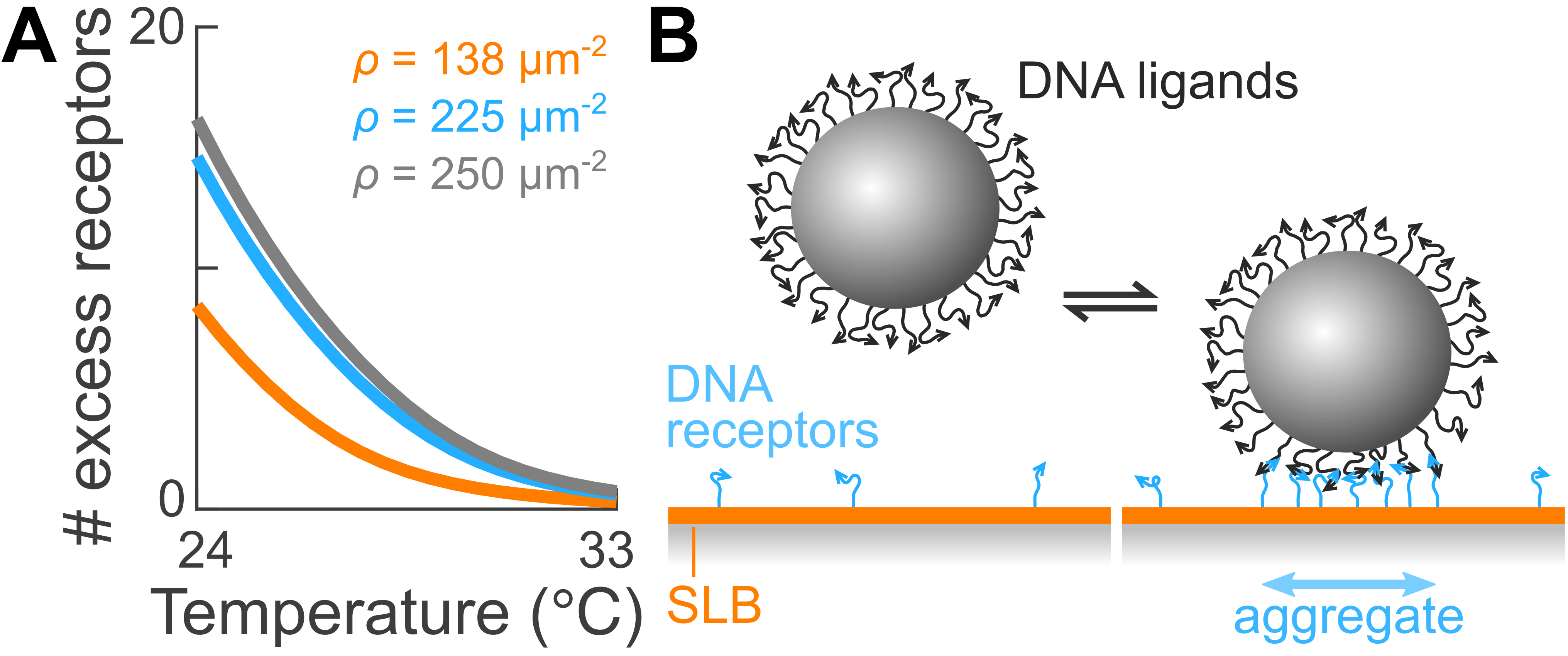}
    \caption{
    Receptor recruitment.
    (A)~Model predictions of the excess number of receptors in the gap between the particle and the supported bilayer as a function of temperature for three receptor densities, showing that receptors are recruited upon binding, and that recruitment is stronger at lower temperatures and larger receptor densities. 
    (B)~Schematics illustrating receptor recruitment.
    }
    \label{fig:fig3}
\end{figure}

Our theoretical model reveals that recruitment of receptors to the binding site---due to their mobility---is responsible for the remaining nonlinearity in the avidity. We confirm this physical picture by computing the excess number of receptors within the gap as a function of the temperature for three receptor densities  (Fig.~\ref{fig:fig3}A).  We find that the number of excess receptors is always positive; thus, mobile receptors are always recruited on average (illustrated in Fig.~\ref{fig:fig3}B). We also find that the excess number of receptors, and thus the extent of recruitment, is larger at lower temperatures. In other words, as the ligand-receptor affinity increases, more receptors can overcome the entropic cost required to enrich the gap between the particle and the membrane. It is precisely this coupling between the ligand-receptor affinity and the entropy penalties of recruiting and confining the receptors in the gap that further enhances the avidity at lower temperatures.
Finally, we note that more receptors are recruited at larger receptor densities and that the number of excess receptors does not plateau at low temperatures, indicating that the ligands grafted to the particles are not saturated by the receptors within the range of receptor densities that we explore. See the ESI$^{\dag}$ for more details on receptor recruitment.

In this section, we showed how the unique combination of DNA as a model ligand-receptor pair, total internal reflection microscopy, and statistical mechanics can shed light on the molecular-scale mechanisms governing adhesion between small particles and membranes. Beyond the effects of multivalency alone, our experiments and model demonstrate that the recruitment of receptors, as well as particle wrapping, play essential roles in determining the avidity of binding, and thus need to be accounted for when designing particles for targeted binding to cell membranes.
These observations constitute one of the first direct experimental validations of the theoretical framework by Mognetti, Frenkel and coworkers.\cite{FrenkelJChemPhys2012, DiMicheleRepProgPhys2019}

\subsection*{Surface mobility}

Whereas the discussion above focused on the thermodynamics of adhesion between colloidal particles and fluid bilayer membranes, targeted delivery could also be influenced by the mobility of adhered particles on the membrane surface, for instance to  collectively remodel the membrane.\cite{DesernoNature2007, CacciutoSoftMatter2013, DinsmoreNanoscale2019} In this section, we determine the relationship between the surface mobility of bound colloids and the physical properties of their receptor aggregates,  as well as the avidity.

We characterize the lateral mobility of membrane-bound particles from their three-dimensional trajectories. First, we segment each trajectory into bound and unbound events  by setting a threshold separation $h_\textrm{b}$ (see the ESI$^{\dag}$ for details and Fig.~\ref{fig:fig1}D for an example). Within each bound event, we then compute the two-dimensional mean squared displacement (MSD) as a function of lag time. Finally, we extract a diffusion coefficient, $D$, for each particle by fitting the average of the mean squared displacement over all bound events to MSD~=~$4Dt$, where $t$ is the lag time.

\begin{figure}[t!]
    \includegraphics{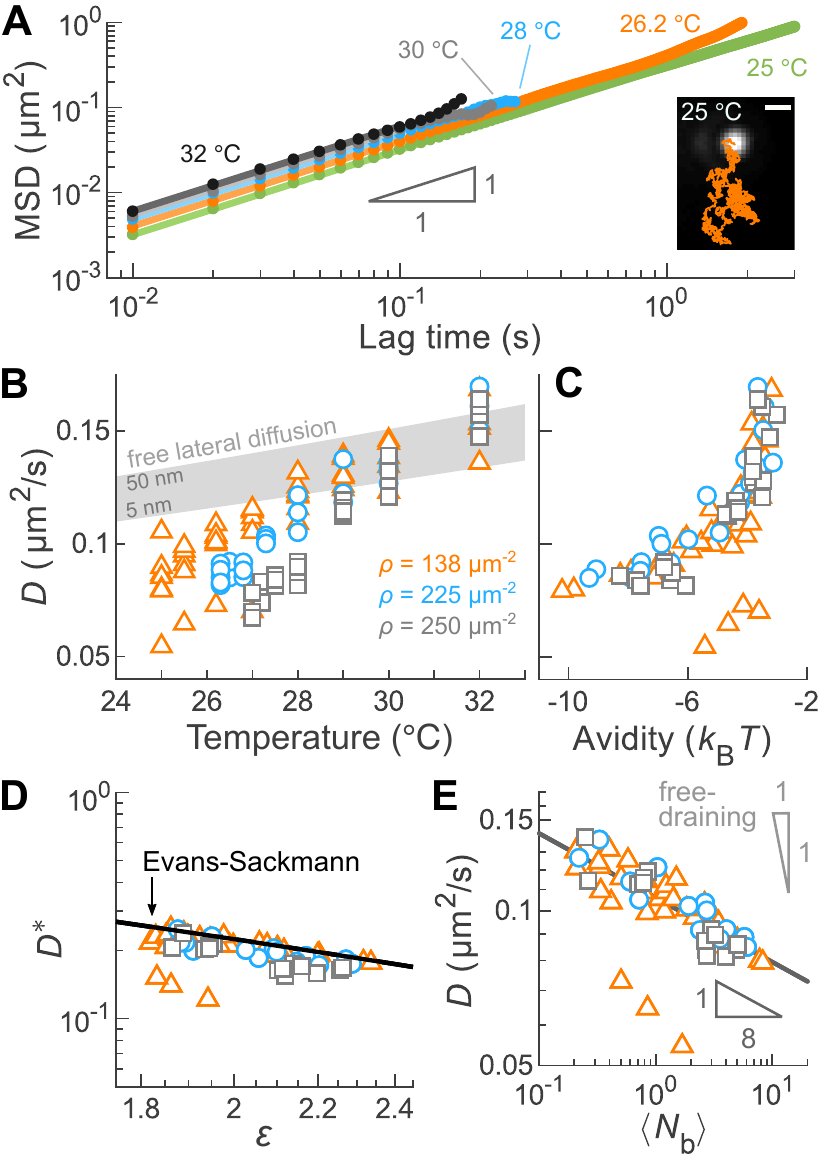}
    \caption{Mobility of membrane-bound particles.
    (A)~Experimental measurements (circles) of the mean-squared displacements (MSD) as a function of lag time, $t$, and their linear fits of $\textrm{MSD} = 4Dt$ (lines), where $D$ is the coefficient of diffusion, for selected temperatures at a fixed density $\rho = 138$~\si{{\micro}m^{-2}}.
    The inset shows a two-dimensional trajectory of a single particle (orange) overlayed on a micrograph of the last frame of the trajectory. Scalebar, 1~\si{{\micro}m}.
    (B)~Measurements of the diffusion coefficient (symbols) of membrane-bound particles as a function of temperature for three receptor densities. The shaded area indicates the range of diffusion coefficients expected for freely diffusing colloids between 5--50~nm above the membrane.\cite{BrennerChemEngSci1961}
    (C)~Measurements of $D$ (symbols) as a function of avidity for three receptor densities collapse to a single curve, demonstrating the coupling between the binding avidity and the latteral mobility.
    (D)~The Evans-Sackmann model \cite{SackmannJFM1988} quantitatively predicts our measurements. $D^\ast$ is a dimensionless diffusion coefficient and $\varepsilon$ is a dimensionless aggregate radius, defined in the main text. Symbols show calculated values of $D^\ast$ and $\varepsilon$ from experimental measurements. The black solid curve shows predictions of the Evans-Sackmann model with no adjustable parameters.
    (E)~In contrast, the free-draining model fails to describe our measurements. Points show measurements of $D$ with respect to the average number of bonds ${\langle N_\text{b} \rangle}$. The data collapses to a power law with an exponent of roughly $-1/8$ (solid line).
    Physical parameters for the models are the membrane viscosity, $\eta_\textrm{m} = 0.14$--0.19~\si{Pa.s}, the bulk viscosity, $\eta_\textrm{b} = 0.80$--0.93~\si{mPa.s}, the membrane thickness, $h_\textrm{m} = 3.8$~nm, and the glass-membrane separation, $H = 3.1$~nm.
    }
    \label{fig:fig4}
\end{figure}

The membrane-bound particles undergo Brownian diffusion and have mobilities that are strongly correlated with both temperature and receptor density. Fig.~\ref{fig:fig4}A shows representative MSDs at the lowest receptor density for a range of temperatures. All MSDs grow linearly with time, indicating Brownian diffusion. The MSDs are truncated at the highest temperatures due to the short bound lifetimes. Fig.~\ref{fig:fig4}B shows the diffusion coefficients for every particle we studied. For a given grafting density, the diffusion coefficient $D$ increases by a factor of roughly two upon increasing temperature over the full range. 
Additionally, the diffusion coefficient $D$ is smaller than---or within the estimated range of---the diffusion coefficient expected for colloids diffusing freely between 5--50~nm above the membrane,\cite{BrennerChemEngSci1961} which suggests that binding hinders surface mobility. 

Furthermore, increasing receptor density decreases the diffusion coefficient $D$ at fixed temperature, suggesting again that adhesion plays a significant role in determining the mobility of membrane-bound particles. 
Indeed, rescaling the diffusion coefficient by the binding avidity collapses all of our experimental measurements to a single curve (Fig.~\ref{fig:fig4}C), suggesting that the avidity is the essential physical variable governing mobility. 

To elucidate the physical origin of this coupling, we
hypothesize that the lateral mobility of membrane-bound particles is dictated by the mobility of their receptor aggregates within the bilayer membrane. To explore the relationship between ligand-receptor binding and surface mobility, we compare our experimental measurements to two classical models: (1)~a model from Evans and Sackmann;\cite{SackmannJFM1988} and (2) the free-draining model.\cite{Flory1953, Kucik1999, FalkeBiophysJ2010, HookNanoLett2016} These two models differ in how they compute the hydrodynamic drag on an inclusion diffusing within a fluid membrane. In our system, we take the inclusion to be the aggregate of cholesterol molecules that tether the receptors to the lipid membrane. The free-draining model assumes that the inclusion is permeable to lipids and unbound receptors, such that the total drag on the inclusion is simply the sum of the drag against each cholesterol anchor, leading to
\begin{equation}
    D \propto {\langle N_\text{b} \rangle}^{-1},
    \label{eq:free-draining}
\end{equation}
where ${\langle N_\text{b} \rangle}$ is the average number of ligand-receptor bonds.\cite{Flory1953, Kucik1999, FalkeBiophysJ2010, HookNanoLett2016} 
In the opposite limit, the inclusion is completely impermeable and diffuses like a single, unit aggregate. Saffman and Delbr{\"u}ck first predicted the diffusion coefficient $D$ from the aggregate radius, $R$, for free membranes.\cite{Saffman1975} This model was later extended by Evans and Sackmann for the case of supported lipid bilayers to account for the hydrodynamic interactions between the membrane constituents---both the lipids and the aggregate---and the support.\cite{SackmannJFM1988} 
The Evans-Sackmann model predicts:
\begin{equation}
     D = \dfrac{k_\textrm{B}T}{4\pi\eta_\textrm{m}h_\textrm{m}}  
     \left[ \dfrac{\varepsilon^2}{4} \left( 1 + \dfrac{b_\textrm{p}}{b_\textrm{s}} \right) + \dfrac{\varepsilon\,K_1(\varepsilon)}{K_0(\varepsilon)} \right],
     \label{eq:mobility-ES}
\end{equation}
where $\eta_\textrm{m}$ is the membrane viscosity, $h_\textrm{m}$ is the membrane thickness, $b_\textrm{p}$ is the inclusion-substrate coefficient of friction, $b_\textrm{s}$ is the membrane-substrate coefficient of friction, and
$K_\nu$ are the modified Bessel functions of the second kind.  $\varepsilon$ is a dimensionless aggregate radius:
\begin{equation}
    \varepsilon \approx R \left(  \dfrac{\eta_\textrm{b}}{H h_\textrm{m}\eta_\textrm{m}} \right)^{1/2},
    \label{eq:ES}
\end{equation}
where $\eta_\textrm{b}$ is the viscosity of the bulk fluid and $H$ is the separation between the membrane and its substrate. This approximate form of $\varepsilon$ is accurate in the limit of $H \ll L_\textrm{SD}$, where $L_\textrm{SD} = h_\textrm{m} \eta_\textrm{m}/ 2 \eta_\textrm{b}$ is the Saffman-Delbr{\"u}ck length giving the range of hydrodynamic coupling between membrane inclusions.\cite{DiamantBiophysJ2009} In our experiments, $H = 3.1$~nm due to the \mbox{PEGylated} lipids and $L_\textrm{SD} = 300$--400~nm; hence the condition $H \ll L_\textrm{SD}$ is met. For simplicity, we assume $b_\textrm{p} = b_\textrm{s}$. See the ESI$^{\dag}$ for more details.

To test these two predictions, we rescale our measurements of the diffusion coefficient using a combination of experimental and theoretical results. Whereas we measure the diffusion coefficient, $D$, directly in our experiments, we cannot measure the average number of bridges in an aggregate, ${\langle N_\text{b} \rangle}$, nor the aggregate radius, $R$. Instead, we rely on predictions from our statistical mechanical model to \emph{infer} these two quantities from our measurements of the well depth of the interaction potentials. Specifically, we create phenomenological relationships between ${\langle N_\text{b} \rangle}$ and the well depth, and between $R$ and the well depth using predictions from our model. We then use these one-to-one mappings to infer the average number of bridges and the aggregate radius from the measured interaction potentials (see the ESI$^{\dag}$).

We find that the Evans-Sackmann model quantitatively describes our measurements of the mobility of membrane-bound particles. Defining a dimensionless diffusion coefficient, $D^\ast = D 4\pi\eta_\textrm{m}h_\textrm{m}/k_\textrm{B}T$, which corresponds to the translational mobility of an inclusion within the membrane, we find that all of our experimental measurements of $D^\ast$ collapse on a single curve when plotted as a function of the dimensionless aggregate radius $\varepsilon$ (Fig.~\ref{fig:fig4}D). Moreover, plugging in the physical constants for our experimental system, we see that the Evans-Sackmann model quantitatively predicts both the trend and the magnitude of the dimensionless diffusion coefficient $D^\ast$. In contrast, the Saffman-Delbr{\"u}ck model \cite{Saffman1975, WhiteJFM1981,PetrovBiophysJ2008} is off by roughly one order of magnitude, as shown in the ESI.$^{\dag}$ In retrospect, this result is unsurprising since the membranes in our experiments are supported on a glass substrate.

The validity of the Evans-Sackmann model is likely due to strong hydrodynamic coupling between the cholesterol anchors within the receptor aggregates. As mentioned above, the range of hydrodynamic coupling between membrane inclusions is given by the Saffman-Delbr{\"u}ck length, $L_\textrm{SD}$, which is roughly 300--400~nm in our experiments. 
By comparison, we estimate that the receptors within the aggregate are separated by only 40--80~nm. Because this estimate is one order of magnitude smaller than the Saffman-Delbr{\"u}ck length, we hypothesize that receptors are strongly coupled.

In contrast, the free-draining model does not reproduce our experimental observations. Replotting our measurements of the diffusion coefficient $D$ against ${\langle N_\text{b} \rangle}$ collapses the data onto a single curve with a power-law exponent of $-1/8$ (Fig.~\ref{fig:fig4}E). While the collapse indicates that ${\langle N_\text{b} \rangle}$ is a relevant parameter in governing the lateral diffusion coefficient $D$, the $-1/8$ dependence that we find is much weaker than the $-1$ prediction of the free-draining model (Eq.~\ref{eq:free-draining}). Interestingly, in a system sharing many similar features as our own, Block and collaborators found that the diffusion of lipid vesicles adhered to a supported bilayer by few, long-lived DNA bonds was accurately described by the free-draining model.\cite{HookNanoLett2016} While we estimate that the typical distances between DNA bonds in their system and ours are comparable, other features---such as the fraction of PEGylated lipid and, possibly, the DNA receptor density---are not. In particular, the receptors in the studies by Block and coworkers were anchored to the membrane using cholesterol molecules, while we use cholesterol-triethylene glycol (TEG) modifications. Because the hydrophilic TEG groups enable the anchors to spontaneously insert deep into the lipid bilayer,\cite{HusterJPhysChemB2009,LiuLangmuir2018} we hypothesize that the details of the hydrodynamic drag differ in the two cases.

\section*{Conclusions}

We set out to elucidate the physical principles that determine the thermodynamics and dynamics of multivalent ligand-receptor binding between small particles and fluid membranes, with the ultimate goal of identifying the key players implicated in targeted drug delivery.  Using a new experimental model system combining DNA-coated colloids with DNA-labeled lipid bilayers, we characterized and modeled adhesion, surface mobility, and their interplay in multivalent ligand-receptor binding. We showed that the strength of adhesion---or avidity---is a strongly nonlinear function of the affinity of the individual ligand-receptor pairs. This nonlinearity results from three contributions: (1) the statistical mechanics of multivalent binding; (2) the recruitment of mobile receptors embedded within the membrane; and (3) the adhesion-mediated elastic deformations of the membrane. We also found that membrane-bound particles undergo two-dimensional Brownian motion, with a mobility that is dictated by that of their aggregate of receptor anchors. Combining theoretical predictions with direct experimental measurements, we demonstrated that the mobility of a membrane-bound particle is accurately predicted by the Evans-Sackmann model of impermeable, solid inclusions diffusing in supported membranes. This result suggests a strong hydrodynamic coupling between the cholesterol molecules within the aggregate. Taken together, our study provides one of the first direct experimental validations of the theoretical framework developed by Mognetti, Frenkel and coworkers \cite{FrenkelJChemPhys2012, DiMicheleRepProgPhys2019} for a system with fixed ligands and mobile receptors---a configuration with direct relevance to targeted drug delivery.\cite{WhitesidesAngewChem1998, KiesslingACSChemBiol2007, FrenkelPNAS2017}

Going forward, our findings suggest that future approaches to designing targeted interactions between colloidal particles and fluid membranes should include the mobility of receptors and the deformability of the membrane, in addition to the specificity of ligand-receptor binding. As we hypothesize in this article, membrane deformations are important because they can occur for relatively weak interactions, yet produce a substantial increase in avidity. This enhancement is due to the large increase in contact area that can be generated by even small deformations. For instance, in our experimental system, we estimate that deformations of roughly 0.5~nm can increase avidity by roughly 1~$k_\textrm{B}T$, an amount that is comparable to the contributions from multivalency or receptor mobility. Therefore, we suggest that future models of targeted binding also determine the shape of the membrane by minimizing the elastic energy considering contributions from membrane bending and stretching,\cite{Helfrich1953} in addition to the adhesion energy.\cite{FrenkelJChemPhys2012, DiMicheleRepProgPhys2019} One such approach was recently implemented in a theoretical study of receptor-mediated endocytosis.\cite{MognettiPRE2018}

Finally, we envision that our results and experimental approaches---mediating interactions using DNA ligand-receptor pairs---could be extended to control and study the self-assembly of colloidal particles bound to lipid vesicles. Self-assembly of nanometer-scale particles bound to fluid membranes, such as membrane proteins, is central to many biological processes, including membrane trafficking, cell division, and cell movement.\cite{McMahonNature2005,UngerCell2008} Furthermore, the cooperative assembly and folding of membranes and membrane-bound proteins can also generate amazing nanostructured materials, like the structurally colored wing scales of many butterflies.\cite{PrumPNAS2010} There, deformations of the membrane give rise to long-range elastic forces between inclusions that direct them to self-assemble. Could we recapitulate similar processes using colloids that bind to and deform membranes? \cite{DufresnePRE2016,KraftSciRep2016} Using DNA to control the self-assembly of colloids on lipid vesicles could open new possibilities in programmable self-assembly. One unique feature of our DNA-based approach is that the adhesion energy can be tuned in situ via the temperature, and predicted using the model validated in this article. Moreover, one can even imagine studying self-assembly of multiple particle species with orthogonal ligand-receptor pairs, different particle sizes, different adhesion strengths, and thereby different degrees of wrapping. We anticipate that such multicomponent systems could produce a complex diversity of structures that far exceeds the types of structures that can be built from colloids or lipids alone.

\section*{Materials and Methods}

\paragraph*{DNA-grafted particles.}
We synthesize 3-(trimethoxysilyl)propyl methacrylate (TPM) colloids using an emulsification technique.\cite{PineNatCommun2015} The synthesized particles are 1.43-\si{{\micro}m}-diameter spheres and have a density of 1.228~\si{\gram\per\cm^3}.\cite{PineSoftMatter2011}
We graft the TPM colloids with dibenzocyclooctyne-amine (DBCON)-modified single-stranded DNA molecules (Integrated DNA Technologies, Inc.) using click chemistry.\cite{PineNatCommun2015}
The particles are stored in aqueous buffer containing 10~mM Tris-HCl/1.0~mM EDTA/pH~=~8.0.

\paragraph*{DNA-grafted supported lipid bilayers.}
We make supported lipid bilayers (SLBs) by fusion of small unilamellar vesicles (SUVs) on a glass coverslip.
This lipid mixture is composed of 97.1\%~(w/w) 1,2-dioleoyl-sn-glycero-3-phosphocholine (18:1 DOPC, Avanti Polar Lipids), 2.4\%~(w/w) 1,2-dioleoyl-sn-glycero-3-phosphoethanolamine-N-[methoxy(polyethylene glycol)-2000] (18:1 PEG2000 PE, Avanti Polar Lipids), and 0.5\%~(w/w) Texas Red ~1,2-dihexadecanoyl-sn-glycero-3-phosphoethanolamine (Texas Red DHPE, Thermo Fisher Scientific).
Briefly, we make SUV suspensions by overnight hydration of a dried lipid film followed by sonication. Since large vesicles scatter visible light while SUVs do not, we visually inspect the suspensions after sonication to make sure that they appear clear. SUV suspensions are stored in aqueous buffer containing 20\%~(v/v) glycerol/10~mM Tris-HCl/1.0~mM EDTA/pH~=~8.0.
We fabricate sample chambers with a combination of glass coverslips (VWR), Parafilm (Bemis Company, Inc.) and polydimethylsiloxane (PDMS, Sylgard~184, Dow). We incubate SUVs with chemically- and plasma-cleaned glass coverslips for 30~min to form the SLB and then wash out excess vesicles.

We functionalize the SLB with DNA receptors using a double cholesterol anchor. To tune the receptor density within the membrane, we adjust the receptor concentration and incubation time. Receptors are formed from two cholesterol-triethylene glycol (TEG)-modified single-stranded DNA molecules (Integrated DNA Technologies, Inc.) by thermal annealing. The short DNA strand is labeled with a 6-FAM fluorophore and the long DNA strand carries the sticky end. Hybridized receptors are stored in aqueous buffer containing 500~mM NaCl/10~mM Tris-HCl/1.0~mM EDTA/pH~=~8.0.

We use a laser scanning confocal microscope (TCS SP8, Leica Microsystems GmbH) equipped with a 20x objective (non-immersion, HCX PL Fluotar, numerical aperture, $\text{NA}=0.50$, Leica Microsystems GmbH) and photomultiplier tubes to visually inspect the SLB and to carry out fluorescence recovery after photobleaching experiments to confirm the mobility of the lipids (Texas Red channel, excitation wavelength 552~nm) and the receptors (6-FAM channel, excitation wavelength 488~nm).

\paragraph*{DNA interactions.}
The DNA ligands and receptors hybridize via complementary sticky ends, 5'-TTTTTT\underline{CTCTTA}-3' and 5'-TTGTCC\underline{TAAGAG}-3', respectively. The underlined portions are the sticky ends which bind to form a 6-basepair duplex. Each sticky end is separated from the base of the strand by a poly-T spacer. We design these DNA sequences so that the particle-membrane binding strength is roughly 1--10~$k_\textrm{B}T$ between 25--35~\si{\degreeCelsius}. Thermodynamic parameters are $\Delta H^\circ = -40.9$~kcal/mol and $\Delta S^\circ = -118.4$~cal/K/mol.\cite{MarkhamNuclAcRes2005,MarkhamChapter2008}

\paragraph*{Interaction potentials and membrane-bound particle mobility.}
We measure the DNA-mediated particle-membrane interactions and the lateral mobility of bound particles using a custom-made, prism-based total internal reflection microscope.  
We match the refractive index of the glass sample chamber to the prism (68\textdegree{}, N-BK7, Tower Optical Corp.) using immersion oil (type N, Nikon Corp.). We control the sample temperature using a thermoelectric module and a thermistor (TE Technology, Inc.) placed under and on top of the prism, respectively, and a custom-made water block.
When a colloidal particle is in the evanescent wave, it scatters an amount of light which decreases exponentially with the particle-glass  separation distance $h$.\cite{Prieve1999} 
Light scattered by the particles is imaged using an upright microscope consisting of a 40x non-immersion objective (infinity-corrected, Plan Fluor, numerical aperture, $\text{NA}=0.75$, Nikon Corp.), a tube lens (focal length, 200~mm, ThorLabs) and a high-speed sCMOS camera (Zyla 5.5, Andor, Oxford Instruments) recording at roughly 100~frames per second.
We measure the scattered intensity as a function of time \cite{Crocker1996} and then construct a histogram of particle-glass separations, $h$, from which we compute the particle-SLB interaction potential by inverting the Boltzmann distribution.
We compute the mean squared displacement of membrane-bound particles during bound events, which we identify using a threshold on the separation $h$.
All experiments were performed in aqueous buffer containing 500~mM NaCl/10~mM Tris-HCl/1.0~mM EDTA/pH~=~8.0.

\paragraph*{Modeling the interactions.}
We use a semi-analytical approach based on the theoretical framework developed by Mognetti, Frenkel and coworkers \cite{FrenkelJChemPhys2012, DiMicheleRepProgPhys2019} to estimate the interactions between a colloidal particle and the membrane. We model the DNA ligands and receptors as ideal chains with 10 and 8 segments, respectively, and of a Kuhn length of $4$~nm. First, we estimate the free energy between a pair of plates separated by distance $\Tilde{h}$. The grafting density of the upper plate matches that of the colloidal particles used in experiment. The lower plate is attached to a grand canonical reservoir to mimic the presence of mobile receptors in our system. The adhesion energy between the two plates is
\begin{equation}
   \beta F_\textrm{adh}(\Tilde{h}) = \rho A(1-\chi_\textrm{r}) - N_\textrm{l}\log \chi_\textrm{l} - N_\textrm{l} \log \, (1 + \overline{N_\textrm{r}} \Xi \chi_\textrm{r}),
    \label{eq_f_finl_si}
\end{equation}
where $\beta = 1/k_\textrm{B}T$, $\rho$ is the density of grand canonical reservoir of receptors, $A$ is the plate area, $N_\textrm{l}$ is the number of ligands, $\overline{N_\textrm{r}}$ is the number of recruited receptors,  $\chi_\textrm{l/r}$ is proportional to the reduction in degrees of freedom associated with confining an ideal chain between two plates, and $\Xi$ is proportional to the confinement-dependent hybridization free energy. We use estimations of $\beta F_\textrm{adh}$, together with the Derjaguin approximation, to estimate the interaction potential between a DNA-grafted spherical  particle and a lipid membrane bearing mobile DNA receptors. More details of our approach can be found in the ESI.$^{\dag}$

\section*{Conflicts of interest}
There are no conflicts to declare.

\section*{Acknowledgements}

We thank Tijana~Ivanovic, Tian~Li, Larry~Friedman, and Thomas~Fai for helpful discussions, Melissa~Rinaldin for her comments on the manuscript, and Francisco~Mello and Gregory~Widberg for the design and fabrication of the cooling block. We acknowledge support from the National Science Foundation (DMR-1710112), the Brandeis MRSEC Bioinspired Soft Materials (DMR-1420382 and DMR-2011846), the National Institutes of Health (R01GM108021 from the National Institute Of General Medical Sciences), and the Smith Family Foundation. We acknowledge computational resources provided by NSF XSEDE (award number MCB090163) and the Brandeis HPCC which is partially supported by DMR-2011846.

%%%END OF MAIN TEXT%%%


\balance

\renewcommand\refname{References}

%%%REFERENCES%%%
\bibliography{references} 
\bibliographystyle{rsc} %the RSC's .bst file



\section*{Supplementary material}

# Electronic Supplementary Information for:

# Avidity and surface mobility in multivalent ligand-receptor binding

Simon Merminod,[1,*] John R. Edison,[2,†] Huang Fang,[1]

Michael F. Hagan,[1] and W. Benjamin Rogers[1,‡]

[1]*Martin A. Fisher School of Physics,*

*Brandeis University, Waltham, MA 02453, USA*

[2]*Brandeis Materials Research Science and Engineering Center,*

*Brandeis University, Waltham, MA 02453, USA*

---

[*] Current address: Department of Molecular and Cellular Biology, Harvard University, Cambridge, MA 02138, USA

[†] Current address: Department of Chemical and Biomolecular Engineering, Johns Hopkins University, Baltimore, MD 21218, USA

[‡] wrogers@brandeis.edu

# CONTENTS

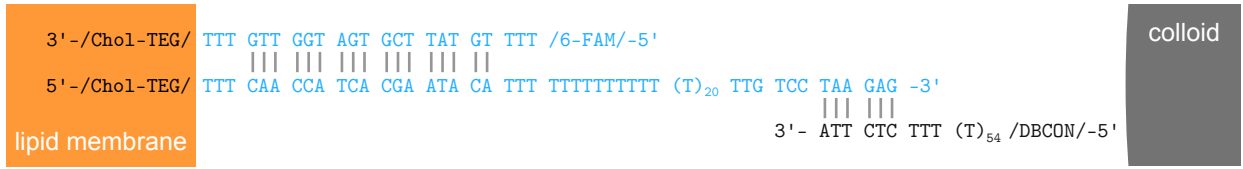

FIG. 1. Schematic of the oligonucleotide sequences of the receptors (left) and the ligands (right). The short vertical grey lines show Watson-Crick base pairing between complementary oligonucleotides.

I. DNA CONSTRUCTS

Transient interactions between a colloidal particle and a supported phospholipid bilayer are due to hybridization of DNA "ligands" and "receptors." We call the DNA molecules grafted to colloids "ligands" and the molecules attached to the lipid bilayer "receptors." The ligands are 63-bases-long, single-stranded, and consist of an inert poly-T spacer and a sticky end on the 3' end (Figure 1). The 5' end is attached to the surface of colloidal particles, as described below (Section II). The sticky end is 6-bases-long and binds the ligand to a receptor. The specific sequences are shown in Figure 1.

The membrane-anchored DNA receptors are complexes formed of two DNA molecules. The short strand is 23-bases-long and has a 6-FAM fluorophore on the 5' end, a TTT spacer, a 17-nucleotides (nt) binding domain, another TTT spacer, and a cholesterol-triethylene glycol (TEG) modification on the 3' end. The long strand is 65-bases-long and consists of a cholesterol-TEG modification on the 5' end, a TTT spacer, a 17-nt binding domain, a 39-bases-long spacer mostly composed of poly-T, and a 6-nt sticky end on the 3' end. The two 17-nt binding domains are complementary to one another. Thus the two receptor molecules hybridize to form the complex shown in Figure 1. The two cholesterol-TEG modifications ensure that the receptors remain bound to the lipid bilayer throughout the duration of our experiment [1]. The fluorophore 6-FAM confirms that the DNA receptors are mobile within the supported bilayer via fluorescence recovery after photobleaching (FRAP).

All three strands are purchased from Integrated DNA Technologies and purified by high-performance liquid chromatography.

3

## II. TPM PARTICLE SYNTHESIS AND LIGAND GRAFTING

We synthesize DNA-grafted colloidal particles made from 3-(trimethoxysilyl)propyl methacrylate (TPM). In brief, we follow a modified version of the method developed by Pine and co-workers [2], which is comprised of three parts: (1) TPM emulsion droplets are polymerized with surface-bound chlorine groups; (2) Chlorine groups on the surface of the TPM particles are substituted with azide groups; and (3) Single-stranded DNA molecules are conjugated to the surface of TPM particles by strain-promoted click chemistry. The specific protocol that we use is described below. Note that sodium azide and ammonium hydroxide are hazardous substances which require specific precautions.

We make chlorine-modified particles by copolymerizing TPM emulsion droplets with 3-chloropropyltrimethoxysilane in five steps. (1) To make TPM emulsion droplets, we add 300 μl of TPM into a 20 ml aqueous solution containing 1% (w/w) ammonium hydroxide and stir for 4 hours at 1000 RPM. (2) We add 30 μl of 3-chloropropyltrimethoxysilane and stir for 30 min. (3) We add 5 ml of 5% (w/w) sodium dodecyl sulfate solution, stir for 10 min, add 7.5 mg of azobis(isobutyronitrile), which initiates polymerization, and stir for 20 min. Then we transfer the solution into an oven at 80 °C for over 4 hours. (4) After polymerization, we wash the TPM particles with a solution containing 0.2% (w/w) Pluronic F-127 four times and resuspend the particles in 10 ml of 0.4% Pluronic F-127 after the final wash.

We substitute the chlorine groups with azide groups to make azide-modified TPM particles. First, we add 10 mg of potassium iodide and 10 ml of 5% sodium azide solution into the particle solution. Then we place the mixture in an oven at 70 °C for 12 hours. After the reaction, we wash the particles with 0.1% (w/w) aqueous Triton X-100 solution and resuspend the particles in 20 ml of 0.1% (w/w) aqueous Triton X-100 solution for storage after the final wash.

Finally, we attach DNA molecules to the azide-modified TPM particles using strain-promoted click chemistry. We mix 317 μl of deionized water, 43 μl of azide-modified TPM particle solution, 20 μl of 100 μM DBCO-modified single-stranded DNA, 40 μl 10x PBS buffer, and place the suspension on a rotator for 24 hours. After the reaction, we wash the DNA-coated particles with 1xTE buffer containing 1% (w/w) Pluronic F-127 five times by centrifugation and resuspension. Then we wash again the particles, with 1xTE five times by centrifugation and resuspension. We store the particles in 1xTE at 4 °C.

## III. SAMPLE PREPARATION

We make supported phospholipid bilayers (SLBs) by fusing small unilamellar vesicles (SUVs) on a glass coverslip. Then we functionalize the SLB with DNA receptors by incubation, followed by washing.

The SLBs are made of a mixture of phospholopids. This mixture is composed of 97.1% (w/w) 1,2-dioleoyl-sn-glycero-3-phosphocholine (18:1 DOPC, Avanti Polar Lipids), 2.4% (w/w) 1,2-dioleoyl-sn-glycero-3-phosphoethanolamine-N-[methoxy(poly-ethylene glycol)-2000] (ammonium salt) (18:1 PEG2000 PE, Avanti Polar Lipids), and 0.5% (w/w) Texas Red 1,2-dihexadecanoyl-sn-glycero-3-phosphoethanolamine, triethylammonium salt (Texas Red DHPE, Thermo Fisher Scientific). All lipids are suspended in chloroform and stored in chloroform at $-20$ °C. Note that chloroform is a hazardous substance which requires specific precautions. PEGylated lipids separate the SLB from the glass coverslip to promote the mobility of membrane-bound objects within the membrane, and to help prevent non-specific binding between colloidal particles and the glass. Texas Red-labeled lipids are used to confirm that the lipids are mobile within the SLB via fluorescence recovery after photobleaching (FRAP).

First, we make an aqueous solution of small unilamellar vesicles. We mix together 1.03 mg of chloroform-suspended lipids in a culture tube at the (w/w) ratios mentioned above. Then we slowly evaporate chloroform to spread the lipids into a dry, thin film at the bottom of the tube. We vacuum desiccate the lipid film for 4 hours to evaporate any remaining chloroform. We hydrate the dried lipid film overnight with 500 µl of 20% glycerol/1xTE hydration buffer to obtain a suspension of large, multilamellar vesicles. The next day, we sonicate the vesicles for 90 minutes to break the multilamellar vesicles into small unilamellar vesicles. Since large vesicles scatter visible light while SUVs do not, we visually inspect the suspension after sonication to make sure that it appears clear. We further dilute the obtained SUV suspension with another 500 µl of hydration buffer to reach a lipid concentration of 1.03 mg/ml in 20% glycerol/1xTE. Finally, we remove any remaining large vesicles by three cycles of centrifugation and dilution. We store the SUV suspension at 4 °C. The final concentration of lipids is 1.03 mg/ml in 20% glycerol/1xTE hydration buffer. We wash all glassware using acetone, ethanol 70%, and ultrapure water. Then we blow the glassware dry with nitrogen and treat it with air plasma.

We make supported lipid bilayers by fusion of SUVs on cleaned glass coverslips. First, we make a custom-made sample chamber using a large glass coverslip and a small glass coverslip, a Parafilm

mask as a spacer, and two PDMS blocks as inlets and outlets. The glass coverslips are washed using acetone, ethanol 70%, and ultrapure water, blown dry with nitrogen, and plama-cleaned. The Parafilm mask is adhered to glass by placing the "sandwich" of glass-Parafilm-glass on a hot plate at 80 °C for roughly 1 minute. The PDMS blocks are designed with a channel at their base, perforated using a hole puncher to create the inlet and outlet, and plasma-bound to the glass coverslip on each side of the Parafilm-glass chamber. Next we seal all interstices with UV-curable optical glue. We fill the sample chamber with SUV suspension diluted to roughly 0.75 mg/ml in 30 mM NaCl/20% glycerol/1xTE and place it on a hot plate at 37 °C for 30 minutes. During this step, the SUVs fuse with the glass substrate to create a supported lipid bilayer. Finally, we wash out excess SUVs using 1 ml of 20% glycerol/1xTE hydration buffer. The buffer is then replaced by washing the chamber a second time with 1 ml of 500 mM NaCl/1xTE.

We functionalize the SLB with DNA receptors by incubation. First, we anneal the DNA receptors by cooling an equimolar solution containing both DNA molecules at concentrations between 5–45 μM in 500 mM NaCl/1xTE from 90 °C to 25 °C at −0.2 °C/min. Next we incubate the SLB with a solution of these FAM-labeled DNA receptors at a concentration between 1–5 μM in 500 mM NaCl/1xTE for 10–60 minutes (Table I). During incubation, pairs of cholesterol molecules incorporate into the SLB [3]. Then we remove any excess receptors by washing with 500 μl of 500 mM NaCl/1xTE. We evaluate the receptor density using confocal fluorescence microscopy and assess their mobility within the SLB using FRAP (TCS SP8, Leica Microsystems GmbH).

Finally, we add DNA-grafted particles at roughly 0.0015% (v/v) in 500 mM NaCl/1xTE and then seal the chamber.

TABLE I. Conditions for functionalization of supported lipid bilayers with DNA receptors.

| Relative receptor density | Receptor suspension concentration (μM) | Incubation time (min) | Best-fit density from the model, $\rho$ (μm$^{-2}$) |
|---|---|---|---|
| low | 1 | 10 | 138 |
| medium | 2.5 | 10 | 225 |
| high | 5 | 60 | 250 |

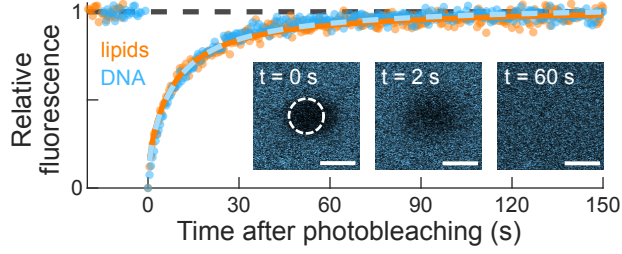

FIG. 2. Relative fluorescence intensity as a function of time after photobleaching for the lipids (orange) and the DNA receptors (blue). Circles are experimental data; solid and dashed curves are model fits using Eq. 3; the horizontal dashed line shows the pre-bleaching fluorescence level. Insets show confocal micrographs of fluorescent DNA receptors after a laser pulse inside the dashed circle. Scalebars, 10 μm.

## IV. FLUORESCENCE RECOVERY AFTER PHOTOBLEACHING

We perform fluorescence recovery after photobleaching experiments to estimate the diffusion coefficients of lipids and membrane-bound DNA receptors within the SLBs. We bleach a 10-μm-diameter spot at the center of a roughly 73 μm × 73 μm frame. The reference region is either: (i) a 10-μm-wide square boundary around the edge of the image; or (ii) a 5-μm-wide octagon surrounding the bleached spot with a 10-μm separation. Pre-bleaching consists of 30 frames at a 500 ms time interval. Photobleaching consists of exposing a species to 100% of the available laser power during a short duration. To achieve strong enough photobleaching, we expose the lipids for 7440 ms (or 40 frames); to bleach the DNA receptors we expose for only 372 ms (1 frame). Post-bleaching consists of 500 frames at a 500 ms time interval. All measurements are carried out at room temperature between 21–23 °C. Figure 2 shows examples of FRAP experiments on lipids and DNA receptors within the same membrane.

We normalize all FRAP curves to account for photobleaching and intensity fluctuations in the imaging laser. Specifically, we compute the normalized fluorescence intensity, $I_{\text{dn}}(t)$, using

$$I_{\text{dn}}(t) = \frac{I_{\text{spot}}(t)}{\langle I_{\text{spot}} \rangle_{\text{pre}}} \frac{\langle I_{\text{ref}} \rangle_{\text{pre}}}{I_{\text{ref}}(t)}, \tag{1}$$

where $t$ is the post-bleaching time, $I_{\text{spot}}(t)$ is the intensity of the bleached spot at time $t$, $\langle I_{\text{spot}} \rangle_{\text{pre}}$ is the average intensity of the bleached spot during the pre-bleaching time, $I_{\text{ref}}(t)$ is the intensity of the reference area at time $t$, and $\langle I_{\text{ref}} \rangle_{\text{pre}}$ is the average intensity of the reference area during

7

pre-bleaching time. Then we compute the fluorescence trace, $I_{\text{fluo}}(t)$, according to

$$I_{\text{fluo}}(t) = \frac{I_{\text{dn}}(t) - I_{\text{dn}}(t=0)}{1 - I_{\text{dn}}(t=0)}. \tag{2}$$

We model the FRAP curves assuming a uniform circular beam and full recovery. More specifically, we fit the fractional fluorescence recovery $f(t)$ to the functional form

$$f(t) = \exp\left(-\frac{2\tau}{t}\right)\left[I_0\left(\frac{2\tau}{t}\right) + I_1\left(\frac{2\tau}{t}\right)\right], \tag{3}$$

where $I_0$ and $I_1$ are modified Bessel functions of first kind at order 0 and 1, respectively, and $\tau$ is the characteristic diffusion time [4]. This functional form is used in Figure 2 to fit the fluorescence recovery of the lipids and the DNA receptors. The diffusion time $\tau$ is related to the diffusion coefficient of the fluorophores, $D_f$, and the radius of the circular beam, $w_b$, by

$$\tau = \frac{w_b^2}{4D_f}. \tag{4}$$

In our FRAP experiments, $w_b = 5$ µm and $D_f$ ranges between 0.8–1.7 µm² s⁻¹ (Table II).

We use a laser scanning confocal microscope (TCS SP8, Leica Microsystems GmbH) equipped with a 20x objective (non-immersion, HCX PL Fluotar, numerical aperture, NA = 0.50, Leica Microsystems GmbH) and photomultiplier tubes. We image and carry out FRAP on the Texas Red-labeled lipids with a laser of wavelength 552 nm and on the FAM-labeled receptors with a laser of wavelength 488 nm.

TABLE II. Diffusion coefficients for lipids and DNA receptors at the three receptor densities of the article.

| Relative receptor density | $D_{\text{lipids}}$ (µm² s⁻¹) | $D_{\text{receptors}}$ (µm² s⁻¹) | Best-fit density from the model, $\rho$ (µm⁻²) |
| --- | --- | --- | --- |
| low | 0.8 | 1.7 | 138 |
| medium | 1.1 | 1.2 | 225 |
| high | 1.1 | 0.9 | 250 |

## V. OPTICAL SETUP

We use a prism-based, total internal reflection microscope to measure the three-dimensional motion of colloidal particles in the vicinity of a supported lipid bilayer. A 671-nm-wavelength

8

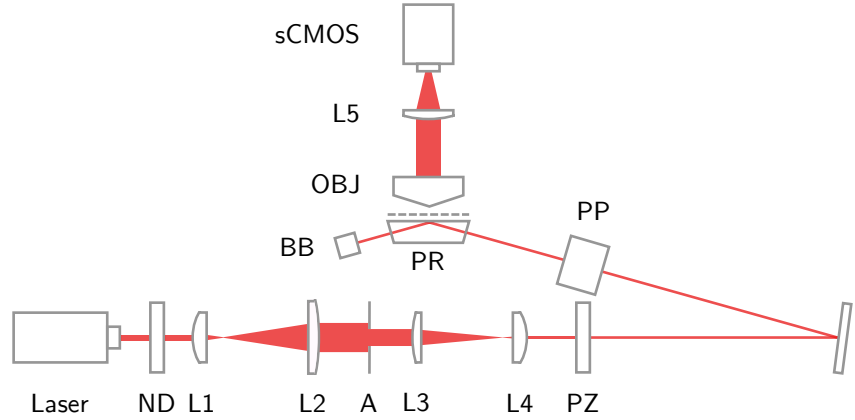

FIG. 3. Optical train of the prism-based total internal reflection microscope. The neutral density filter (ND) adjusts light intensity to prevent saturation in the recorded images while maximizing duration of exposure. The inverted Keplerian telescope made of a pair of spherical, plano-convex lenses (L1 and L2) magnifies the laser beam roughly 8.3x to overfill a 20-mm aperture (A) so as to keep the central part of the Gaussian beam only. The Keplerian telescope made of a pair of spherical, plano-convex lenses (L3 and L4) shrinks the laser beam 10x. The polarizer (PZ) p-polarizes the laser beam before it enters the prism pair (PP) which transforms the beam from circular to elliptical. The elliptical laser beam enters the dove prism (PR) at a right angle and encounters the glass-aqueous sample interface at a 68° angle, where it roughly spreads into a disk. The laser beam generates an evanescent wave inside the sample (dashed line) and is reflected out of the prism into a beam block (BB). The upright microscope composed of an infinity-corrected, 40x objective (OBJ), a spherical, plano-convex lens (L5), and a high-speed sCMOS camera (sCMOS), images the light scattered by colloidal particles from inside the sample.

laser beam (300 mW, SDL-671-LN-300T, Shanghai Dream Lasers) is totally internally reflected at the glass-water interface to create an evanescent wave (Figure 3). Light scattered by colloids diffusing in the evanescent wave is imaged using an upright microscope, consisting of a infinity-corrected 40x objective (Plan Fluor, numerical aperture, NA = 0.75, Nikon Corp.), a tube lens ($f_5$ = 200 mm, spherical, plano-convex lens, ThorLabs) and a high-speed sCMOS camera (Zyla 5.5, Andor, Oxford Instruments) placed in the image plane of the tube lens. We control the temperature of the sample using a thermoelectric module (TE Technology, Inc.) attached to the prism and a thermistor (TE Technology, Inc.) placed on the sample. The thermoelectric module is cooled by a custom-made water-cooling block.

We shape the imaging laser upstream from the total internal reflection microscope to create a

uniform intensity across the field of view. We first magnify the original beam using a 8.3x inverted Keplerian telescope ($f_1$ = 60 mm, $f_2$ = 500 mm, spherical, plano-convex lenses, ThorLabs), of which we crop the outer part of the beam using a 20-mm aperture (ThorLabs). The circular beam is then reduced by a 10x Keplerian telescope ($f_3$ = 500 mm, $f_4$ = 50 mm, spherical, plano-convex lenses, ThorLabs), before being p-polarized by a linear polarizer (Thorlabs). We next transform the circular beam into an elliptical beam using an anamorphic prism pair (magnification, 4.0x, Thorlabs) to create a roughly circular spot upon reflection. A Littrow dispersion prism (N-BK7, Edmund Optics) attached on the exit side of a 68-degree dove prism (N-BK7, Tower Optical Corp.) guides the reflected laser beam out of the setup to a beam block. We adjust the beam intensity using a neutral density filter (optical density, OD = 1, Thorlabs). We align the beam using the back reflection from the dove prism-air interface.

## VI. CALIBRATION OF THE TOTAL INTERNAL REFLECTION MICROSCOPE

We calibrate the relationship between the scattered intensity and the separation distance using the separation-dependent hydrodynamic interactions between a colloidal particle and a flat wall. We assume that the relationship between the scattered intensity, $I$, and the separation between the glass substrate and the bottom of the particle, $h$, is given by

$$I(h) = I_0 \, e^{-h/h_0}, \tag{5}$$

where $I_0$ is the intensity of light scattered by a particle in contact with the glass substrate at $h = 0$ and $h_0$ is the typical penetration depth of the evanescent wave [5]. Then we determine the parameters $I_0$ and $h_0$ using the method from Volpe, Bechinger and co-workers [6], which relies on longstanding hydrodynamic theories of diffusion of a single colloidal particle near a wall [7]. Briefly, we find the parameters $I_0$ and $h_0$ that best match the measured distributions of frame-to-frame particle displacements to theoretical predictions over a range of separations between $h$ = 100–300 nm, where there are no DNA interactions. We consider this calibration process to be successful because we obtain a satisfying match between experimental measurements of the transverse component of the diffusion coefficient, $D_{\perp,\mathrm{exp}}(h)$ [8], and theoretical predictions, $D_{\perp,\mathrm{th}}(h)$ [7] (Figure 4).

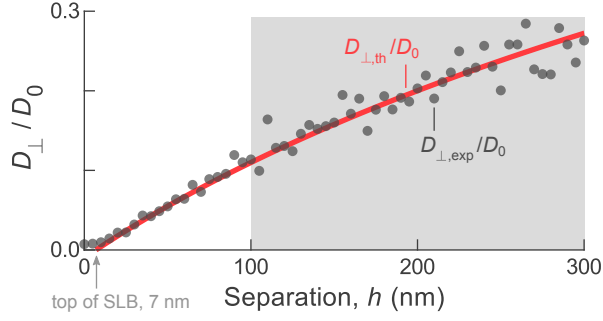

FIG. 4. Example of validation of our calibration routine. Coefficients of transverse diffusion normalized by the Stokes-Einstein coefficient of diffusion, $D_0$, as a function of particle-glass separation, for an experiment after determining $I_0$ and $h_0$. Circles show experimental measurements; the solid red curve shows the theoretical prediction. We assume a non-slip boundary condition at the upper surface of the SLB. We fit $I_0$ and $h_0$ within the shaded region, so as to avoid separations at which DNA strands interact.

## VII. INTERACTION POTENTIALS

We compute the effective interaction potential between a particle and the supported lipid bilayer at a given temperature by inverting the Boltzmann distribution. Specifically, we compute the histogram of the full separation time-series of a single particle, $P(h)$, and then convert it into a free energy profile, $\Delta F(h)$, by inverting the Boltzmann distribution, $P(h) \propto \exp\left[-\Delta F(h)/k_B T\right]$, where $k_B$ is the Boltzmann constant and $T$ is the temperature. This free energy contains two main contributions: one from DNA-mediated interactions and another due to gravity. We subtract the gravitational contribution to $\Delta F$ to obtain the specific DNA-mediated interaction potentials. For each particle, we take the gravitational potential to be the best fit of a straight line through the interaction potential at separations between 100–300 nm, where we are confident that there are no DNA interactions.

We compare our measured interaction potentials to blurred versions of our theoretical potentials. First, we compute the corresponding distribution of separations for a given model potential using the Boltzmann distribution. Then, to simulate the finite precision of our measurements of the separation, we convolve the distribution with a Gaussian kernel of standard deviation $\sigma_h$. This operation smooths the distribution. Then we convert the convolved distribution back into a blurred potential by reinverting the Boltzmann distribution. In practice, this operation makes the potential well shallower and wider than that of the original simulated potential.

11

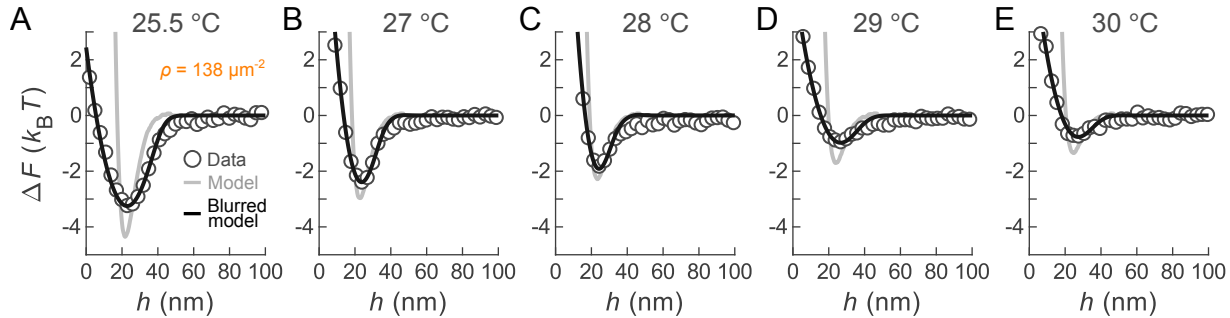

FIG. 5. Colloid-membrane interaction potentials. Interaction potentials from experiments (circles) and numerical simulations (solid curves) for $T$ = 25.5 °C (A), 27 °C (B), 28 °C (C), 29 °C (D), and 30 °C (E), with the receptor density $\rho$ = 138 μm$^{-2}$. The grey curves show the simulated potentials. The black curves show the corresponding blurred potentials with best-fit temperatures constrained within ±0.25 °C of the measured $T$, and $\sigma_h$ = 6.4 nm (A), 3.9 nm (B), 3.0 nm (C), 6.7 nm (D), and 6.0 nm (E). The experimental potentials have been horizontally lined up with the blurred potentials.

Second, we determine the best-fit blurred potentials. We hold the receptor density $\rho$ fixed at the values provided in the main text, and allow the temperature $T$ and our measurement precision $\sigma_h$ to vary. We take the best-fit blurred potential to be the one that minimizes the sum of squared differences with the experimental potential. We also allow the separation $h$ to freely vary to account for errors in our calibration of $I_0$ and $h_0$, with shifts of roughly 0–20 nm. The values that we find for the best-fit temperatures typically vary within ±1.5 °C from the measured temperatures, $T$, which is due in part to the variability in the binding strength between particles. The values of $\sigma_h$ vary between 4–9 nm, which is consistent with measurements of particles bound irreversibly to the coverslip.

We find that the blurred potentials match our experimentally measured potentials. Figure 5 shows experimental potentials, theoretical potentials, and the corresponding blurred potentials at selected temperatures for the receptor density $\rho$ = 138 μm$^{-2}$. The best-fit temperatures are constrained within ±0.25 °C of the measured temperature $T$. In Figure 5, as well as in Figure 2A–C in the article, experimental potentials were shifted as described above.

## VIII. DEFINITION OF THE AVIDITY

We compute the avidity, $\Delta G_{\text{av}}$, from the continuous interaction potential, $\Delta F(h)$. Our definition of avidity follows the definition in Ben-Tal *et al.* [9]; it can also be found in Ref. [10]. We define avidity from the integral of the partition function over the *bound state*, which we define as the range of separation distances over which the particle and receptors interact significantly:

$$\Delta G_{\text{av}} = -k_\text{B}T \, \log\left[(c_\circ N_\text{A})^{1/3} \int_0^{\lambda_\text{b}} e^{-\Delta F(h)/k_\text{B}T} \text{d}h\right], \quad (6)$$

where $c_\circ = 1$ mol/l is a reference concentration, $N_\text{A}$ is Avogadro's number, and $\lambda_\text{b}$ is a binding length scale, or the maximum separation at which ligands can bind to receptors. Since the value of $\Delta G_{\text{av}}$ changes with $\lambda_\text{b}$, even for separations $h$ at which $\Delta F(h) = 0$, it is important to choose $\lambda_\text{b}$ carefully. In our system, we find that $\lambda_\text{b} = 34$ nm yields avidities $\Delta G_{\text{av}}$ that are reasonably insensitive to the precise value of $\lambda_\text{b}$. Thus we set $\lambda_\text{b} = 34$ nm for calculations. Note that this measure of avidity—unlike the well-depth of the interaction potential for example—allows one to compare potentials that have different widths or functional forms. Additionally, because we use a Gaussian kernel to obtain the blurred potentials from the theoretical potentials, a given theoretical potential and the blurred potentials obtained from it have the same avidity, whereas the well-depth alone is a strong function of blurring.

With this definition of avidity, even situations with no interaction, $\Delta F(h) = 0$, will contribute to the avidity since there is a non-zero probability of observing such a separation. In contrast, separations with a large free energy $\Delta F(h) \gg 1$, such as small separations where the repulsion dominates, will not contribute to the avidity.

## IX. AVIDITY MEASUREMENTS FOR PLOT OF AVIDITY V. TEMPERATURE (FIGURE 2D)

To compute the avidity from the experimental potentials—which are discrete and potentially noisy—we fit them with blurred simulated potentials before evaluating Equation (6). Specifically, for each experimental potential, we find the best-fit blurred potential as described in the previous section, and then compute the avidity from the fitted potential using Equation (6). The obtained avidities constitute the data points in Figure 2D of the article.

Next, for each experimental data set at a given receptor density, we identify the receptor density, $\rho$, that provides the best fit between theoretical and experimental avidities. Specifically, we compute the avidity as a function of temperature from theoretical potentials for a wide range of $\rho$. Then we

compare the theoretical avidity at a fixed $\rho$ to the experimental avidity at a fixed receptor density. The best-fit receptor density, $\rho$, is the one that minimizes the sum of the squared differences between the experimental and theoretical avidities across all temperatures. We find that the experimental data at low receptor density is best fit by simulations at $\rho = 138$ μm$^{-2}$, the medium receptor density data by $\rho = 225$ μm$^{-2}$, and the high receptor density data by $\rho = 250$ μm$^{-2}$.

## X. THEORY OF MULTIVALENT LIGAND-RECEPTOR BINDING

We use a recently developed statistical mechanical theory of multivalent interactions to predict the interaction potential between a solid colloid and flat fluid membrane [11]. We begin by discussing the interaction between two flat plates decorated with complementary DNA strands separated by distance $h$. We then use this information together with the Derjaguin approximation to estimate the interaction between the colloid with fixed tethers and an infinite membrane with mobile tethers. Plate 1 of our system has $N_l = \rho_l A$ fixed ligands, where $\rho_l$ is the grafting density of the ligands on the colloid and $A$ is the area of the plate. Plate 2 has mobile receptors and is in contact with a grand canonical reservoir of receptors. When the two surfaces do not interact or when the separation distance $h \to \infty$, the density of receptors equals that of the grand canonical reservoir, which we simply denote $\rho$. As described by Mognetti, Frenkel and coworkers, [11, 12] we write the adhesion free energy between the two plates as

$$\beta F_{\text{adh}}(h) = -\log \sum_{N_r, N_{lr}} Z_{\text{adh}}(h, N_r, N_{lr}), \tag{7}$$

where $Z_{\text{adh}}(h, N_r, N_{lr})$ gives the weight of microstates that contain $N_r$ receptors and $N_{lr}$ bonds, and $\beta = 1/k_B T$ is the inverse thermal energy. Note that $N_{lr}$ is identical to $N_b$ in the article and Section XIV. We assume that the linkers are ideal chains, which allows us to describe $Z_{\text{adh}}(h, N_r, N_{lr})$ as a product of three terms: $Z^l_{\text{conf}}(h, N_l)$, $Z^r_{\text{conf}}(h, N_r)$, and $Z_{\text{bind}}(h, N_r, N_{lr})$. The first two terms correspond to the cost of confining the chains between two plates and the third term gives the likelihood of forming $N_{lr}$ bonds from $N_l$ ligands and $N_r$ receptors:

$$Z_{\text{adh}}(h, N_r, N_{lr}) = Z^l_{\text{conf}}(h, N_l) \sum_{N_r} Z^r_{\text{conf}}(h, N_r) \sum_{N_{lr}} Z_{\text{bind}}(h, N_r, N_{lr}). \tag{8}$$

Using the saddle-point approximation, the sum on the right is approximated to be equal to its dominant term. The dominant term corresponds to $N_{lr} = \overline{N_{lr}}$ and $N_r = \overline{N_r}$, and is found by equating

14

the first derivative of $Z_{\text{adh}}$ with respect to $N_r$ and $N_{lr}$ to zero. Note that $\overline{N_{lr}}$ and $\overline{N_r}$ are separation $h$ dependent. The adhesion free energy can then be written as

$$\beta F_{\text{adh}}(h) = -\log Z^l_{\text{conf}}(h, N_l) - \log Z^r_{\text{conf}}(h, \overline{N_r}) - Z_{\text{bind}}(h, \overline{N_r}, \overline{N_{lr}}) \quad (9)$$

$$= \beta F^l_{\text{conf}}(h) + \beta F^r_{\text{conf}}(h) + \beta F_{\text{bind}}(h). \quad (10)$$

The term $F^l_{\text{conf}}(h) = -k_B T \log Z^l_{\text{conf}}(h)$ is the entropic cost of confining the ligands between two plates separated by distance $h$,

$$\beta F^l_{\text{conf}}(h) = -N_l \log \chi_l(h), \quad (11)$$

with

$$\chi_l(h) = Q_1(h)/Q_1(\infty), \quad (12)$$

where $Q_1(h)$ is the partition function of a ligand tethered at one end and free at the other end confined between two plates separated by distance $h$. In the next section, we describe how to compute $\chi_l(h)$ using Monte-Carlo simulations. In a similar manner for the mobile receptors, $F^r_{\text{conf}}(h)$ can be written as

$$\beta F^r_{\text{conf}}(h) = -\overline{N_r} \log \chi_r(h) - \overline{N_r} \log \frac{\rho A}{\overline{N_r}} + (\rho A - \overline{N_r}). \quad (13)$$

The second term in the above Equation (13) is the chemical potential cost of having $\overline{N_r}$ ideal receptors in the system, which is in equilibrium with a grand-canonical reservoir of density $\rho$.

The partition function $Z_{\text{bind}}$ is computed by considering all possible states that have $N_{lr}$ bonds out of $N_r$ receptors and $N_l$ ligands:

$$Z_{\text{bind}}(h, N_r, N_{lr}) = \binom{N_l}{N_{lr}}\binom{N_r}{N_{lr}} N_{lr}! \, \Xi^{N_{lr}}. \quad (14)$$

Each bond has an associated statistical weight $\Xi$ which is proportional to the binding affinity of a ligand-receptor pair and equals $e^{-\Delta G^\circ} \langle e^{-\Delta G_{\text{conf}}(h)} \rangle$. Here $\Delta G^\circ$ is the temperature-dependent free-energy change associated with hybridization of the complementary base pairs, and $\Delta G_{\text{conf}}$ is the confinement-strength-dependent entropic cost associated with connecting the free ends of a ligand and a receptor. Since the receptors are mobile, for a given separation $h$, we average $\Delta G_{\text{conf}}$ over all possible tether positions of the receptor. Therefore we enclose this term in angled brackets.

15

As mentioned earlier, using the saddle point approximation on Equation (8) we obtain the free energy change associated with ligand-receptor binding as

$$\beta F_{\text{bind}}(h) = N_l \ln \frac{N_l - \overline{N_{lr}}}{N_l} + \overline{N_r} \ln \frac{\overline{N_r} - \overline{N_{lr}}}{\overline{N_r}} + \overline{N_{lr}}, \tag{15}$$

where the average number of receptors $\overline{N_r}$ and bonds $\overline{N_{lr}}$ are given by

$$\overline{N_{lr}} = \frac{N_l \rho A \Xi \chi_r}{1 + \rho A \Xi \chi_r} \tag{16}$$

$$\overline{N_r} = \overline{N_{lr}} + (\rho A \chi_r). \tag{17}$$

Putting all the terms together, we obtain

$$\beta F_{\text{adh}}(h) = \rho A (1 - \chi_r) - N_l \log \chi_l - N_l \log (1 + \overline{N_r} \Xi \chi_r). \tag{18}$$

In the next section, we describe our model system specifically and how we use Monte Carlo simulations to compute $\chi_{l/r}$ and $\Xi$.

## XI. MODEL AND NUMERICAL SCHEME

We first compute the interactions between two flat plates separated by a distance $h$ and then use the Derjaguin approximation to estimate the interaction between the colloid and the flat membrane. The lower plate is attached to a grand-canonical reservoir of receptors and the upper plate has ligands anchored at fixed positions. We model the ligands as ideal chains with 10 segments each. The mobile receptors are also ideal chains, and each consists of 8 segments. The length of each segment is given by the Kuhn length of single-stranded DNA, which is 4 nm in 500 mM NaCl [13], as in our experiments. The first segment of the receptor equals 5.8 nm, as it is anchored to the membrane via a short double-stranded DNA domain. All lengths are made dimensionless by the Kuhn length and energies by $k_B T$. The interaction between a pair of plates with receptor (reservoir) density $\rho$ and ligand grafting density $\rho_l$ is given by Equation (10). Below we describe how we obtain the various terms involved in Equation (18). We employ the simulation techniques suggested in Appendix A of Varilly *et al.* [12]. Details of these simulations can also be found in Reference [14].

Determining the conformational free-energy cost of confining ligands $\beta F^1_{\text{conf}}(h)$ requires an estimate of $\chi_l(h)$, which is the ratio of the partition function of an unbound ligand between two

plates at separation $h$ to the partition function at infinite separation $h \to \infty$, $Q(h)/Q(\infty)$. To obtain this quantity, we use Rosenbluth sampling. In this technique, we grow the chain segment-by-segment. For each segment, we consider $k$ different trial directions chosen from the surface of a sphere with radius equal to the segment length and compute a weight associated with each trial. The weight $w_i$ equals 0 or 1 depending on whether the segment overlaps with the walls or not. We randomly pick one of the $k$ trial segments based on the weight $w_i$. The overall Rosenbluth weight of the chain is given by $W = \prod_i^{N_{\text{seg}}} w_i/k$. Here $N_{\text{seg}}$ is the total number of segments in the chain, which equals 8 for the receptors and 10 for the ligands. The average Rosenbluth factor, $\langle W \rangle$, gives $Q(h)/Q_1$, where $Q_1 = \prod_i^{N_{\text{seg}}} 4\pi l_i^2$ is the partition function of an ideal chain in bulk and $l_i$ is the segment length. The factor $\chi_l(h)$ is obtained from $\langle W \rangle(h)/\langle W \rangle(\infty)$. The conformational cost of confining the receptors $\chi_r(h)$, is computed in the same manner.

The attractive part of the interaction between the plates is given by $\beta F_{\text{bind}}(h)$. To obtain an estimate of this term, we need to compute $\Xi = e^{-\Delta G°} \langle e^{-\Delta G_{\text{conf}}(h)} \rangle$, with $\Delta G° = \Delta H° - T \Delta S°$ as the binding affinity between the sticky ends of the DNA ligands and receptors. For the sequences we use in the experiments $\Delta H° = -40.9$ kcal/mol and $\Delta S° = -0.1184$ kcal/mol K. As mentioned earlier, $\Delta G_{\text{conf}}$ is the confinement-strength-dependent entropic cost associated with turning a ligand and a receptor, each initially with one free end, into a single chain that is bound at both ends to the plates. This quantity depends on the distance between the ligand and the receptor $|\mathbf{r}_l - \mathbf{r}_r|$, and the plate separation $h$. Since the receptors are mobile, we average this quantity over all possible tether positions of the receptor [15]. For fixed $\{|\mathbf{r}_l - \mathbf{r}_r|, h\}$ we have

$$\Delta G_{\text{conf}}(h) = \frac{p(N_{\text{seg}}, |\mathbf{r}_l - \mathbf{r}_r|)}{\rho_0} \frac{\langle W_{\text{lr}} \rangle}{\langle W_l \rangle \langle W_r \rangle}, \qquad (19)$$

where $\rho_0$ is the standard concentration, $p(N_{\text{seg}}, |\mathbf{r}_l - \mathbf{r}_r|)$ is the probability of having a chain with $N_{\text{seg}}$ begin at $|\mathbf{r}_l|$ and end at $|\mathbf{r}_r|$, or the probability that a chain with $N_{\text{seg}}$ has end-to-end distance $|\mathbf{r}_l - \mathbf{r}_r|$, and $\langle W_l \rangle$, $\langle W_r \rangle$, and $\langle W_{\text{lr}} \rangle$ are defined below. For the case where all the segments are of equal length, exact analytical expressions are available to compute $p$. Since one of our segments in the receptor is slightly larger than the others, we numerically estimate [16] the probability $p$ from the equation

$$p(N_{\text{seg}}, |\mathbf{r}_l - \mathbf{r}_r|) = \frac{1}{2\pi^2 |\mathbf{r}_l - \mathbf{r}_r| l_0} \int_0^\infty \frac{\sin(|\mathbf{r}_l - \mathbf{r}_r| x) \sin(l_0 x) \sin^{N_{\text{seg}}-1}(x)}{x^{N_{\text{seg}}-1}} dx \qquad (20)$$

where $l_0$ is the length of the first segment of the receptors, which in reduced units equals 1.45. $\langle W_l \rangle$ and $\langle W_r \rangle$ in Equation (19) are average Rosenbluth weights of the unhybridized ligand and

receptor confined between two plates separated by distance $h$. To obtain the Rosenbluth weight of the hybridized strand $\langle W_{\text{lr}} \rangle$, which is tethered at both ends, we need to modify the Rosenbluth scheme [12]. In the case where the strand is tethered only at one end, we propose the location of new trial segments uniformly on the surface of the sphere. However for the hybridized strand, since both ends are tethered, we need to weigh each trial segment $i$ with the probability to reach the final tether point in $N_{\text{seg}} - i$ steps. This is done using rejection sampling. More details on this scheme can be found in Reference [17].

Once we obtain $\langle \Delta G_{\text{conf}} \rangle$, we can estimate $\Xi$ and use equations (17),(15) and (10) to estimate the interaction between two plates separated by $h$. This data is used together with the Derjaguin approximation to compute the interaction free-energy between a colloidal particle and the membrane.

## XII. RECEPTOR RECRUITMENT

Due to their mobility within the fluid lipid bilayer, receptors can be recruited in the gap between a particle and the membrane to form an aggregate. Figure 6A shows the excess number of receptors in the gap as a function of the temperature and separation distance for a fixed receptor density, $\rho = 138 \; \mu m^{-2}$. Note that this figure uses similar data to Figure 3A in the article, but shows the dependence on separation distance instead of using a Boltzmann-averaged value, which Figure 3*A* did. As in the article, the data shows that mobile receptors are always recruited on average, and that recruitment is larger at lower temperatures. Additionally, here we observe that receptor recruitment is a nonmonotonic function of the colloid-membrane separation. More specifically, we find a temperature-dependent separation distance at which receptor recruitment is maximal. Qualitatively, this optimum separates the separation distances at which the entropy loss due to vertical confinement is weak compared to the enthalpy gain due to forming new bonds—the large separations—from the separations at which vertical confinement is strong—the small separations. For a more detailed physical picture on the effect of vertical confinement on receptor recruitment, let us now examine the receptor density profiles.

To strike a balance between the enthalpy gain and the entropy loss, the receptors self-organize at the binding site. Figure 6B shows radial density profiles of mobile receptors relative to the density of their grand canonical reservoir, $\rho = 138 \; \mu m^{-2}$, for selected separations at a fixed temperature $T = 28 \; °C$. The spatial distribution of mobile receptors is nonuniform, in sharp

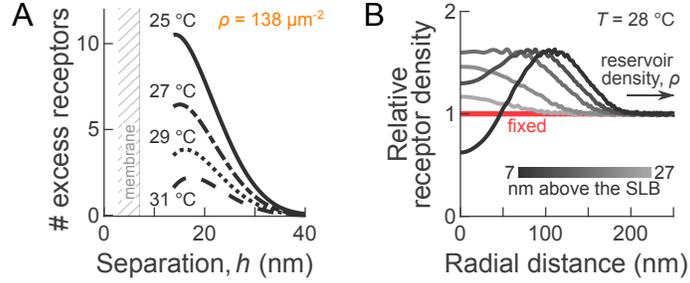

FIG. 6. Receptor recruitment. (A) Model predictions of the excess number of receptors in the gap between the particle and the supported bilayer at selected temperatures for a fixed receptor density $\rho = 138\ \mu m^{-2}$, showing that receptors are recruited into an aggregate upon binding. (B) Radial density profiles of the receptors relative to the density of the grand canonical reservoir, $\rho = 138\ \mu m^{-2}$, for different separations: 7–27 nm (black to light gray). The receptors organize at the periphery of the binding site at small separations. The red line shows the uniform density profile for fixed receptors.

contrast with the fixed receptors, which are uniformly distributed on the membrane. Furthermore, the shape of the mobile receptor distribution varies with colloid-membrane separation and reveals that receptors self-organize to minimize their entropic cost of confinement while being recruited. At large separations, the highest receptor density occurs at the center of the recruited aggregate. In contrast, at small separations, the receptors accumulate at the periphery of the binding site to maximize their conformational entropy while still binding to ligands. Both of these features are roughly independent of the receptor density $\rho$.

## XIII. WRAPPING OF AN ADHERING PARTICLE BY THE LIPID MEMBRANE

Lipid membranes can undergo elastic deformations. In particular, a membrane can wrap around an adhering particle, either partially or totally. Deserno [18] theoretically predicted the regions of the relevant parameter space where such wrapping occurs, and by what amount. The relevant parameters are the membrane bending rigidity, $\kappa$, the membrane tension, $\sigma$, the adhering particle radius, $a$, and the particle-membrane adhesion energy per unit area, $w$. Deserno defines two dimensionless ratios: $\tilde{w} = 2wa^2/\kappa$ and $\tilde{\sigma} = \sigma a^2/\kappa$. He predicts no membrane deformation for $\tilde{w} < 4$, partial wrapping for $4 < \tilde{w} < 4 + 2\tilde{\sigma}$, and full wrapping for $\tilde{w} > 4 + 2\tilde{\sigma}$. For our DOPC membranes, typical values are $\sigma = 1$ pN nm$^{-1}$ and $\kappa = 20\ k_B T$ [19, 20]. We take the well-depth of the plate-plate interaction free-energy per unit area computed in the simulations as the adhesion

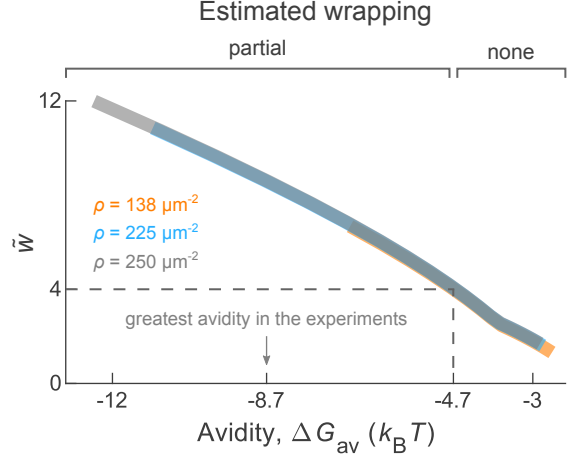

FIG. 7. Estimated wrapping of a colloid adhering to an elastic lipid membrane. The values of dimensionless parameter $\tilde{w}$ tell us whether the membrane is deformed and is computed here for the receptor densities and range of temperature explored in the experiments: $\rho$ = 138 μm$^{-2}$, 225 μm$^{-2}$ and 250 μm$^{-2}$, and $T$ = 25–32 °C. The membrane is predicted to not deform when $\tilde{w} < 4$ and to partially wrap the particle when $\tilde{w} > 4$.

energy, $w$. We find that $w$ ranges between 30–240 $k_\text{B}T/\mu\text{m}^2$ for $\rho$ between 138–250 μm$^{-2}$ and temperatures between 25–32 °C. Therefore, the dimensionless ratio $\tilde{w}$ ranges between roughly 2–12 and $\tilde{\sigma} \approx 6000$.

We find that wrapping occurs when the avidity is below $-4.7\ k_\text{B}T$. Figure 7 shows $\tilde{w}$ as a function of the avidity from our simulations at the three receptor densities $\rho$ = 138 μm$^{-2}$, 225 μm$^{-2}$, and 250 μm$^{-2}$, and temperatures between 25–32 °C. All of the data collapses on a single curve, for which the condition $\tilde{w} > 4$ corresponds to avidities $\Delta G_\text{av} < -4.7\ k_\text{B}T$.

## XIV. MOBILITY OF A MEMBRANE-BOUND PARTICLE

### A. Identifying the bound state

We determine whether or not a particle is bound based on a separation threshold, $h_\text{b}$. For each particle, we determine $h_\text{b}$ from the blurred potential $\beta\Delta F$ that has the same well-depth as the measured potential. Specifically, we take $h_\text{b}$ to be the separation distance at which $\beta\Delta F(h_\text{b}) = -0.1$. We define $h_\text{b}$ as a threshold between the bound state and the unbound state. Next, we define bound events as segments of trajectories for which the separation $h(t)$ is below the threshold $h_\text{b}$. All

20

separations above $h_b$ are considered to be unbound.

## B. Bound lifetimes statistics

We define a bound event as an uninterrupted segment of a trajectory during which the particle is always bound. For each of these segments, we measure the lifetime of the bound event. Figure 8 shows the distribution of these bound lifetimes for three different temperatures. For each temperature, we see a broad distribution of lifetimes. The means of the distributions depend on temperature: The particles stay bound for a longer duration on average at low temperatures as compared to high temperatures. This observation agrees with expectations, since we expect the average bound lifetimes to increase with increasing adhesion strength.

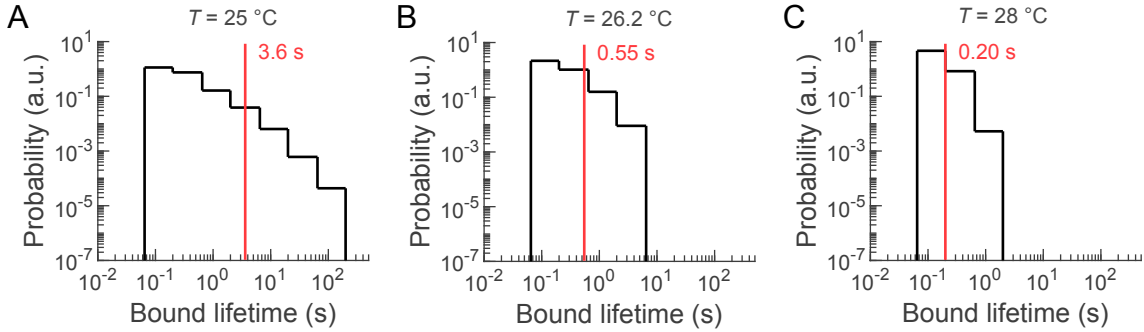

FIG. 8. Bound lifetime distributions at (A) $T = 25$ °C, (B) $T = 26.2$ °C, and (C) $T = 28$ °C for the lowest receptor density, $\rho = 138$ μm$^{-2}$. The vertical red lines indicate the mean value of each distribution. The ranges of $h_b$ are 41–48 nm (A), 42–45 nm (B), and 38–45 nm (C).

## C. Relationships between potential well depth, receptor aggregate properties and avidity

We compute phenomenological relationships between the average number of bonds, $\langle N_b \rangle$, and the average receptor aggregate radius, $R$, from our model predictions of the interaction potential $\Delta F(h)$. Specifically, we compute $\langle N_b \rangle$ as the Boltzmann-average over the number of bonds formed at each separation, $N_b(h)$:

$$\langle N_b \rangle = \frac{\int_0^\infty N_b(h)\, e^{-\beta \Delta F(h)} \, dh}{\int_0^\infty e^{-\beta \Delta F(h)} \, dh}. \tag{21}$$

21

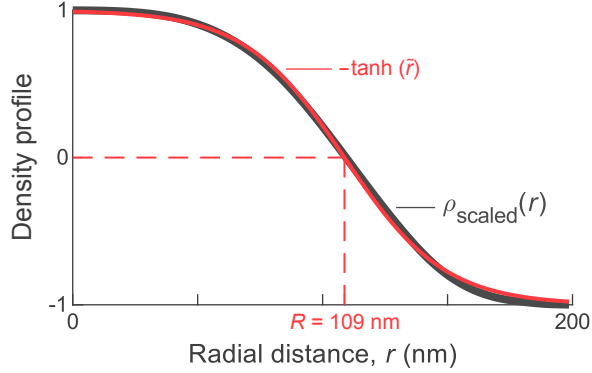

FIG. 9. Estimating the radius of the receptor aggregate from the density profile. Representative rescaled receptor density profile, $\rho_{\text{scaled}}(r)$, as a function of the radial distance to the point of closest contact, $r$, for receptor reservoir density $\rho = 138\ \mu\text{m}^{-2}$ and $T = 25\ °\text{C}$ (grey curve). The red curve is the best-fit hyperbolic tangent, $-\tanh(\tilde{r})$, with $\tilde{r} = a_r r + b_r$, where $a_r$ and $b_r$ are adjustable parameters. We define the average aggregate radius, $R$, as the radial distance at which $\tanh(\tilde{r})$ reaches its half-height. In this example, $R = 109$ nm.

Similarly, we compute the Boltzmann-averaged receptor density profiles, $\langle \rho_{\text{prof}} \rangle(r)$:

$$\langle \rho_{\text{prof}} \rangle(r) = \frac{\int_0^\infty \rho_{\text{prof}}(r, h)\, e^{-\beta \Delta F(h)}\mathrm{d}h}{\int_0^\infty e^{-\beta \Delta F(h)}\mathrm{d}h}, \tag{22}$$

where $\rho_{\text{prof}}(r, h)$ is the receptor density profile at separation $h$. Finally, we compute normalized receptor profiles $\rho_{\text{scaled}}(r)$:

$$\rho_{\text{scaled}}(r) = 2\, \frac{\langle \rho_{\text{prof}} \rangle(r)/\langle \rho_{\text{prof}} \rangle(\infty) - 1}{\max[\langle \rho_{\text{prof}} \rangle(r)/\langle \rho_{\text{prof}} \rangle(\infty)] - 1} - 1. \tag{23}$$

From each such rescaled profile, we define the average aggregate radius, $R$, as the radial coordinate at which the density profile $\rho_{\text{scaled}}(r)$ reaches half its maximum height (Figure 9). In practice, we identify the aggregate radius by fitting a hyperbolic tangent of the form $-\tanh(\tilde{r})$, with $\tilde{r} = a_r r + b_r$, where $a_r$ and $b_r$ are adjustable parameters.

Our model predictions of the average number of bonds per aggregate and the average aggregate radius collapse onto universal curves. Figure 10A,B show $\langle N_b \rangle$ and $R$ as a function of the minimum of the interaction potential, $\min \Delta F$, for the receptor densities and temperature ranges discussed in the article. We find that predictions of $\langle N_b \rangle$ for all three densities collapse onto a single curve (Figure 10A). This curve is similar to earlier models of multivalent binding, which predicted that $\min \Delta F$ is proportional to the average number of bridges [21, 22]. Similarly, predictions

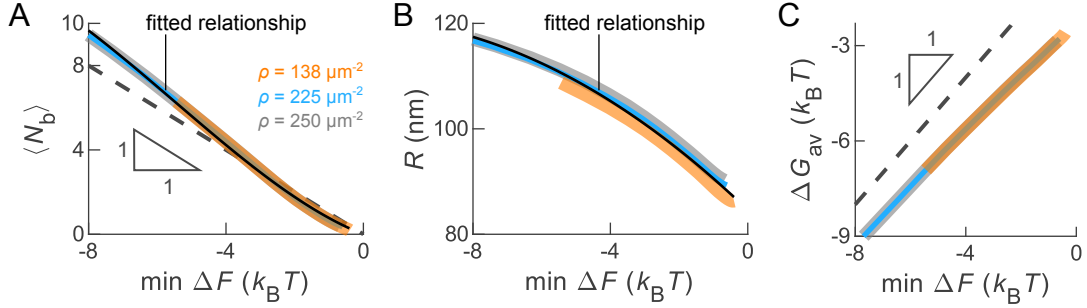

FIG. 10. Relationships between the well depth, the number of bridges, the aggregate radius, and the avidity. The thick colored curves are simulation results for $\rho = 138$ μm$^{-2}$ (orange), 225 μm$^{-2}$ (blue) and 250 μm$^{-2}$ (grey). (A) The average number of bonds is a slightly nonlinear function of the well depth. The dashed line shows $\langle N_b \rangle = -\min(\Delta F / k_B T)$. The black solid curve is a fitted polynomial relationship. The average number of bonds behaves similarly to earlier models [21, 22]. (B) The average aggregate radius increases with the well depth. The black solid curve is a fitted polynomial relationship. (C) The avidity is a roughly linear function of the well depth. The dashed line shows $\Delta G_{av} = \min \Delta F$.

of $R$ collapse onto a single curve for all three densities (Figure 10B). Although we do not have closed-form analytic expressions for $\langle N_b \rangle$ or $R$, we can fit phenomenological expressions to the two curves. Most importantly, these two expressions provide a link between an experimental observable—namely the well depth—and two microscopic parameters—the average number of bridges and the aggregate radius—that we use to test model predictions of the lateral mobility of inclusions within a membrane. In practice, we subtract the average well depth of the nonspecific interaction potentials—roughly 1 $k_B T$—from the well depth of each experimental potential before we use the phenomenological expressions shown in Figure 10A,B to infer values of $\langle N_b \rangle$ and $R$. Figure 10C shows the roughly linear relationship between the avidity and the well depth.

### D. Diffusion of permeable membrane inclusions

Consider an inclusion composed of an aggregate of many identical units diffusing within a lipid membrane. If the lipid molecules in the membrane can freely drain between the units, then the units are *not* coupled to one another hydrodynamically. In this limit, the total drag on the inclusion is just the sum of the drag on the motion of each unit. If the units are identical, the total drag is proportional to the number of units. Combining these assumptions with the Stokes-Einstein relationship, the diffusion coefficient of an inclusion should scale as the inverse of the average

23

number of units: $D_{\text{free-draining}} \propto \langle N_b \rangle^{-1}$. This prediction is often referred to as the *free-draining* model [23–26].

### E. Diffusion of unit-aggregate inclusions

Now consider a different limit, in which the units composing the inclusion behave as a unit-aggregate that *cannot* be penetrated by lipid molecules within the membrane. In this case, the inclusion can be viewed as a non-permeable cylinder diffusing within the membrane. We denote the cylinder radius by $R$.

#### 1. In free membranes

Saffman and Delbrück calculated the translational diffusion coefficient of a unit-aggregate inclusion within a free membrane [27]:

$$\beta D_{\text{SD}} = \frac{1}{4\pi \eta_m h_m} \left[ \ln \frac{2}{\varepsilon'} - \gamma \right], \quad (24)$$

where $\eta_m$ is the two-dimensional dynamic viscosity of the membrane, $\eta_{3D}$ is the bulk dynamic viscosity of the aqueous buffer, $h_m$ is the membrane thickness, $\gamma = 0.5772$ is the Euler constant, and

$$\varepsilon' = \frac{R}{h_m} \frac{2\eta_{3D}}{\eta_m} \quad (25)$$

is a dimensionless inclusion radius. The Saffman-Delbrück model is relevant if $\varepsilon' \leq 0.1$ [28]. In our experiment, $\varepsilon' = 0.23$–$0.30$, thus we cannot use the Saffman-Delbrück model directly.

Hughes, Pailthorpe and White (HPW) derived a generalization of Equation (24) for arbitrary $\varepsilon'$ [28]. To avoid the complicated numerical computations involved, we use the second-order approximation of the HPW results multiplied by the Petrov-Schwille correction [29], which gives an accurate approximation for $\varepsilon' < 10^3$:

$$\beta D_{\text{SDHPW-PS}} = \frac{1}{4\pi \eta_m h_m} \left[ \ln \frac{2}{\varepsilon'} - \gamma + \frac{4\varepsilon'}{\pi} - \frac{\varepsilon'^2}{2} \ln \frac{2}{\varepsilon'} \right] \times \left[ 1 - \frac{\varepsilon'^3}{\pi} \ln \frac{2}{\varepsilon'} + \frac{c_1 \varepsilon'^{b_1}}{1 + c_2 \varepsilon'^{b_2}} \right]^{-1}, \quad (26)$$

where $c_1 = 0.73761$, $b_1 = 2.74819$, $c_2 = 0.52119$, and $b_2 = 0.61465$.

We find that models describing the diffusion of unit-aggregate inclusions within free membranes do not describe our experimental measurements. Figure 11A shows all of our experimental measurements of the dimensionless diffusion coefficient $D^*_{\text{SDHPW-PS}} = 4\pi \eta_m h_m \beta D_{\text{SDHPW-PS}}$ plotted as

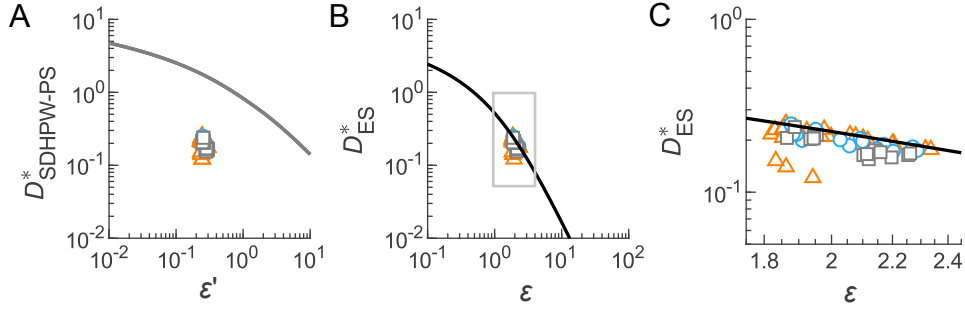

FIG. 11. Mobility of membrane-bound particles. The symbols are all of our experimental measurements. The curves show the models with no adjustable parameters. (A) The model by Hughes, Pailthorpe, White and others [27–29] describing the diffusion of unit-aggregate inclusions in *free* membranes, $D^*_{\text{SDHPW-PS}}(\varepsilon') = 4\pi\eta_m h_m \beta D_{\text{SDHPW-PS}}(\varepsilon')$, from Equation (26), fails to predict the experimental measurements. (B) In contrast, a model by Evans and Sackmann [30] predicting the diffusion of unit-aggregate inclusions in *supported* membranes, $D^*_{\text{ES}}(\varepsilon) = 4\pi\eta_m h_m \beta D_{\text{ES}}(\varepsilon)$, from Equation (27), quantitatively describes both the magnitude and the trend of our data. (C) Enlarged boxed region from panel (B), identical to Figure 4C in the article. $D^*_{\text{ES}}$ is identical to $D^*$ in the article.

a function of the dimensionless aggregate radius $\varepsilon'$. All values of $D^*_{\text{SDHPW-PS}}$ are grouped together, but are roughly one order of magnitude smaller than the model predictions. Thus this model fails to predict the translational mobility of the membrane-bound particles in our experiments. We hypothesize that this disagreement between experiment and theory arises due to hydrodynamic coupling between the receptor aggregate and the glass substrate.

2. *In supported membranes*

Evans and Sackmann calculated the translational diffusion coefficient of a unit-aggregate inclusion within a supported membrane [30]:

$$\beta D_{\text{ES}} = \frac{1}{4\pi\eta_m h_m} \left[ \frac{\varepsilon^2}{4}\left(1 + \frac{b_p}{b_s}\right) + \frac{\varepsilon K_1(\varepsilon)}{K_0(\varepsilon)} \right], \quad (27)$$

where $b_s$ is the membrane-substrate coefficient of friction, $b_p$ is the inclusion-substrate coefficient of friction, $K_\nu$ is the modified Bessel function of the second kind at order $\nu$, and

$$\varepsilon = R\left(\frac{b_s}{\eta_m h_m}\right)^{1/2} \quad (28)$$

25

is a dimensionless aggregate radius. This model accounts for the drag on the inclusion due to the substrate, instead of simply the drag on the inclusion due to the membrane, as in the model for free membranes detailed above.

We estimate the parameters in Equation (27) as follows. To start with, we estimate the size of the gap between the lower leaflet of the bilayer membrane and the glass substrate, which we denote $H$. Indeed if $H$ is small enough, the expression of $\varepsilon$ in Equation (28) can be replaced by a more convenient expression independent from $b_s$.

On the membranes that we make, PEG is in the brush regime with a brush thickness of approximately 3.1 nm. Qualitatively, the transition between the mushroom regime and the brush regime for a polymer species grafted on a surface occurs when the grafted density is large enough that it becomes entropically favorable for the polymer molecules to favor extending transversely from the surface. For polymers grafted on lipids of a bilayer, this transition occurs when the fraction of polymer-grafted lipids, $X_{\text{poly}}$, reaches the threshold

$$X_{\text{poly}}^{\text{m}\to\text{b}} = \frac{A_l}{\pi a_m^2} n_{\text{poly}}^{-6/5}, \qquad (29)$$

with $A_l$ the membrane surface area per lipid molecule, $a_m$ the size of a monomer unit, and $n_{\text{poly}}$ the degree of polymerization [31]. In our experiments, DOPC lipids have each a surface area $A_l = 0.68$ nm$^2$ [32], one oxyethylene monomer unit of PEG is of length $a_m = 0.39$ nm [31], and the degree of polymerization of PEG(2k) is $n_{\text{PEG}} = 45$, thus the threshold fraction of PEG is $X_{\text{PEG}}^{\text{m}\to\text{b}} = 0.015$. We can now compare this threshold value with the fraction of PEGylated lipids in our experiments, which is $X_{\text{PEG}} = 0.024$ (see Section III). We find that $X_{\text{PEG}} > X_{\text{PEG}}^{\text{m}\to\text{b}}$, implying that PEG is in the brush regime. With this information, we can estimate the equilibrium length of the PEG chains on our bilayers [31]:

$$L_{\text{PEG}} \approx n_{\text{PEG}} a_m^{5/3} \left(\frac{X_{\text{PEG}}}{A_l}\right)^{1/3}, \qquad (30)$$

which roughly equals 3.1 nm. Now assuming that this theoretical equilibrium length sets the separation between the PEGylated bilayer and the glass substrate, $H$, we take $H = 3.1$ nm.

The small value of $H$ makes the lubrication approximation valid and thus enables us to use an approximate expression of $\varepsilon$. The lubrication approximation,

$$b_s \approx \eta_{3D}/H, \qquad (31)$$

is valid when the thickness of the layer between the membrane and the substrate, $H$, is much smaller than the characteristic length $R/\varepsilon'$ [30]. In our experiments, $R/\varepsilon' \approx 340$–$390$ nm, thus $H \ll R/\varepsilon'$.

26

We can therefore use Equation (31) to rewrite $\varepsilon$ as

$$\varepsilon \approx \left(\frac{\varepsilon' R}{2H}\right)^{1/2}, \qquad (32)$$

which we use in Figure 4C in the article and here in Figure 11B,C.

We use the following values for the other physical constants: $\eta_\mathrm{m}$, the membrane two-dimensional dynamical viscosity, on which the correction due to $< 1\%$ cholesterol is negligible [33], ranges between 0.14–0.19 Pa s depending on temperature [34]; $\eta_\mathrm{3D}$, the bulk dynamical viscosity of the aqueous buffer with 500 mM NaCl, ranges between 0.80–0.93 mPa s depending on temperature [35]; $h_\mathrm{m}$, the membrane thickness, is 3.8 nm [32, 36]; and we assume $b_\mathrm{s} = b_\mathrm{p}$.

---



\end{document}